\documentclass[11pt]{article}

\usepackage{amsfonts,amsmath}
\usepackage{algorithm2e,framed}
\usepackage{graphicx}
\usepackage{amssymb}

\newcommand{\Expect}[1]{\mbox{}{\bf{E}}\left[#1\right]}

\newcommand{\FNorm }[1]{\mbox{}\left\|#1\right\|_F  }
\newcommand{\FNormS}[1]{\mbox{}\left\|#1\right\|_F^2}

\newcommand{\TNorm }[1]{\mbox{}\left\|#1\right\|_2  }

\newcommand{\XNorm }[1]{\mbox{}\left\|#1\right\|_{\xi}  }

\newcommand{\VTTNorm }[1]{\mbox{}\left|#1\right|_2  }
\newcommand{\VTTNormS}[1]{\mbox{}\left|#1\right|_2^2}

\newcommand{\setlinespacing}[1]%
           {\setlength{\baselineskip}{#1 \defbaselineskip}}

\newcommand{\abs }[1]{\left|#1\right|}

\newtheorem{definition}{Definition}

\newtheorem{lemma}{Lemma}
\newtheorem{theorem}{Theorem}

\newenvironment{Proof}{\noindent {\em Proof:}}{\\\hspace*{\fill}\mbox{$\diamond$}}

\long\def\killtext#1{}

\def\Prob{\hbox{\bf{Pr}}}

\def\diag{\hbox{\bf{diag}}}

\newlength{\defbaselineskip}
\begin{document}

\begin{titlepage}

\title{
Relative-Error CUR Matrix Decompositions
\thanks{
A preliminary version of this paper appeared in manuscript and technical 
report format as ``Polynomial Time Algorithm for Column-Row-Based 
Relative-Error Low-Rank Matrix Approximation''
\cite{DMM06_relerr_110305,DMM06_relerr_TR}.
Preliminary versions of parts of this paper have also appeared as
conference proceedings~\cite{DMM06,DMM06b,DMM06c}.
}
}

\author{
Petros Drineas
\thanks{
Department of Computer Science,
Rensselaer Polytechnic Institute,
Troy, New York 12180,
drinep@cs.rpi.edu.
}
\and Michael W. Mahoney
\thanks{
Yahoo! Research,
Sunnyvale, California 94089,
mahoney@yahoo-inc.com.
Part of this work was performed while at the
Department of Mathematics,
Yale University,
New Haven, Connecticut 06520.
}
\and S. Muthukrishnan
\thanks{
Google, Inc.,
New York, NY,
muthu@google.com.
Part of this work was performed while at the
Department of Computer Science,
Rutgers University,
New Brunswick, New Jersey 08854.
}
}

\date{}
\maketitle


\begin{abstract}
Many data analysis applications deal with large matrices and involve
approximating the matrix using a small number of ``components.''
Typically, these components are linear combinations of the rows and columns
of the matrix, and are thus difficult to interpret in terms of the original
features of the input data.
In this paper, we propose and study matrix approximations that are explicitly
expressed in terms of a small number of columns and/or rows of the data
matrix, and thereby more amenable to interpretation in terms of the original
data.

Our main algorithmic results are two randomized algorithms which
take as input an $m \times n$ matrix $A$ and a rank parameter $k$.
In our first algorithm, $C$ is chosen, and we let $A'=CC^+A$,
where $C^+$ is the Moore-Penrose generalized inverse of $C$. In
our second algorithm $C$, $U$, $R$ are chosen, and we let
$A'=CUR$.
($C$ and $R$ are matrices that consist of actual columns and rows,
respectively, of $A$, and $U$ is a generalized inverse of their intersection.)
For each algorithm, we show that with probability at least $1-\delta$
\[
\FNorm{A-A'} \leq (1+\epsilon) \FNorm{A-A_k},
\]
where $A_k$ is the ``best'' rank-$k$ approximation provided by
truncating the singular value decomposition (SVD) of $A$, and
where $\FNorm{X}$ is the Frobenius norm of the matrix $X$. The
number of columns of $C$ and rows of $R$ is a low-degree
polynomial in $k$, $1/\epsilon$, and $\log(1/\delta)$.
Both the Numerical Linear Algebra community and the Theoretical Computer
Science community have studied variants of these matrix decompositions over
the last ten years.
However, our two algorithms are the first polynomial time algorithms for such
low-rank matrix approximations that come with relative-error guarantees;
previously, in some cases, it was not even known whether such matrix
decompositions exist.
Both of our algorithms are simple and they take time of the order needed to
approximately compute the top $k$ singular vectors of $A$.

The technical crux of our analysis is a novel, intuitive sampling method we
introduce in this paper called ``subspace sampling.''
In subspace sampling, the sampling probabilities depend on the Euclidean norms
of the rows of the top singular vectors.
This allows us to obtain provable relative-error guarantees by deconvoluting
``subspace'' information and ``size-of-$A$'' information in the input matrix.
This technique is likely to be useful for other matrix approximation and data
analysis problems.
\end{abstract}

\end{titlepage}
\newpage
\section{Introduction}
\label{sxn:intro}

Large $m \times n$ matrices are common in applications since the data often 
consist of $m$ objects, each of which is described by $n$ features.
Examples of object--feature pairs include:
documents and words contained in those documents;
genomes and environmental conditions under which gene responses are measured;
stocks and their associated temporal resolution;
hyperspectral images and frequency resolution;
and web groups and individual users.
In each of these application areas, practitioners spend vast amounts of time 
analyzing the data in order to understand, interpret, and ultimately use this 
data for some application-specific task. 

Say that $A$ is the $m \times n$ data matrix.
In many cases, an important step in data analysis is to construct a compressed 
representation of $A$ that may be easier to analyze and interpret.
The most common such representation is obtained by truncating the Singular 
Value Decomposition (SVD) at some 
number $k \ll \min \{m,n\}$ terms, in large part because this provides the 
``best'' rank-$k$ approximation to $A$ when measured with respect to any 
unitarily invariant matrix norm.
Unfortunately, the basis vectors (the so-called eigencolumns and eigenrows) 
provided by this approximation (and with respect to which every column and row 
of the original data matrix is expressed) are notoriously difficult to 
interpret in terms of the underlying data and processes generating that data.
For example, the vector 
[$(1/2)$ {\rm age} - $(1/\sqrt{2})$ {\rm height} + $(1/2)$ {\rm income}], 
being one of the significant uncorrelated ``factors'' from a dataset of 
people's features is not particularly informative.
It would be highly preferable to have a 
low-rank approximation that is nearly as good as that provided by the SVD
but that is expressed in terms of a small number of {\em actual columns} 
and/or {\em actual rows} of a matrix, rather than linear combinations of 
those columns and rows. 

The main contribution of this paper is to provide such decompositions.
In particular, we provide what we call a relative-error {\em CUR matrix
decomposition}: given an $m \times n$ matrix $A$, we decompose it as a product
of three matrices, $C$, $U$, and $R$, where $C$ consists of a small number of
actual columns of $A$, $R$ consists of a small number of actual rows of $A$,
and $U$ is a small carefully constructed matrix that guarantees that the
product $CUR$ is ``close'' to $A$.
In fact, $CUR$ will be nearly as good as the best low-rank approximation to
$A$ that is traditionally used and that is obtained by truncating the SVD.
Hence, the columns of $A$ that are included in $C$, as well as the rows of $A$
that are included in $R$, can be used in place of the eigencolumns and
eigenrows, with the added benefit of improved interpretability in terms of the
original data.

Before describing applications of our main results in the next subsection, we
would like to emphasize that two research communities, the Numerical Linear
Algebra (NLA) community and the Theoretical Computer Science (TCS) community,
have provided significant practical and theoretical motivation for studying
variants of these matrix decompositions over the last ten years.
In Section~\ref{sxn:previous}, we provide a detailed treatment of relevant
prior work in both the NLA and the TCS literature.
The two algorithms presented in this paper are the first polynomial time
algorithms for such low-rank matrix approximations that come with
relative-error guarantees; previously, in some cases, it was not even known
whether such matrix decompositions exist.

\subsection{Applications}
\label{sxn:intro:data_apps}

As an example of this preference for having the data matrix expressed in terms 
of a small number of actual columns and rows of the original matrix, as 
opposed to a small number of eigencolumns and eigenrows, consider recent data 
analysis work in DNA microarray and DNA Single Nucleotide Polymorphism (SNP) 
analysis \cite{KPS02,LA04,Paschou07a}.
DNA SNP data are often modeled as an $m \times n$ matrix $A$, where $m$ is the 
number of individuals in the study, $n$ is the number of SNPs being analyzed, 
and $A_{ij}$ is an encoding of the $j$-th SNP value for the $i$-th individual.
Similarly, for DNA microarray data, $m$ is the number of genes under 
consideration, $n$ is the number of arrays or environmental conditions, and 
$A_{ij}$ is the absolute or relative expression level of the $i$-th gene in 
the $j$-th environmental condition.
Biologists typically have an understanding of a single gene that they 
fail to have about a linear combination of $6000$ genes (and also similarly for 
SNPs, individuals, and arrays);
thus, recent work in genetics on DNA microarray and 
DNA SNP data has focused on heuristics to extract \emph{actual} genes, 
environmental conditions, individuals, and SNPs from the eigengenes, 
eigenconditions, eigenpeople, and eigenSNPs computed from the original data 
matrices \cite{KPS02,LA04}.%
\footnote{For example, in their review article ``Vector algebra in the 
analysis of genome-wide expression data''~\cite{KPS02}, which appeared in 
\emph{Genome Biology}, Kuruvilla, Park, and Schreiber describe many uses of 
the vectors provided by the SVD and PCA in DNA microarray analysis.
The three biologists then conclude by stating that:
``While very efficient basis vectors, the vectors
themselves are completely artificial and do not correspond
to actual (DNA expression) profiles.
...
Thus, it would be interesting to try to find basis
vectors for all experiment vectors, using actual experiment
vectors and not artificial bases that offer little insight.''
That is, they explicitly state that they would like decompositions of the form 
we provide in this paper!}
Our CUR matrix decomposition is a direct formulation of this 
problem: determine a small number of actual SNPs that serve as a basis with 
which to express the remaining SNPs, and a small number of individuals to 
serve as a basis with which to express the remaining individuals.
In fact, motivated in part by this, we have successfully applied a variant of 
the CUR matrix decomposition presented in this paper to intra- and 
inter-population genotype reconstruction from tagging {SNP}s in DNA SNP 
data from a geographically-diverse set of populations \cite{Paschou07a}.
In addition, we have applied a different variant of our CUR matrix 
decomposition to hyperspectrally-resolved medical imaging 
data \cite{MMD06}.
In this application, a column corresponds to an image at a single physical 
frequency and a row corresponds to a single spectrally-resolved pixel, and
we have shown that data reconstruction and classification tasks can be 
performed with little loss in quality even after substantial data 
compression \cite{MMD06}.

A quite different motivation for low-rank matrix approximations expressed in 
terms of a small number of columns and/or rows of the original matrix is to 
decompose efficiently large low-rank matrices that possess additional structure 
such as sparsity or non-negativity.  
This often arises in the analysis of, e.g., large term-document
matrices~\cite{Ste99,Ste04_TR,BPS04_TR}.
Another motivation comes from statistical learning theory, where the data need 
not even be elements in a vector space, and thus expressing the Gram matrix in 
terms of a small number of actual data points is of 
interest \cite{WS01,WRST02_TR,dm_kernel_CONF,dm_kernel_JRNL}.
This procedure has been shown empirically to perform well for approximate 
Gaussian process classification and regression \cite{WS01}, to approximate the 
solution of spectral partitioning for image and video 
segmentation \cite{FBCM04}, and to extend the eigenfunctions of a 
data-dependent kernel to new data points \cite{BPVDRO03,LafonTH}.
Yet another motivation is provided by integral equation applications 
\cite{GE96,GTZ97,GT01}, where large coefficient matrices arise that have 
blocks corresponding to regions where the kernel is smooth and that are thus
well-approximated by low-rank matrices.
In these applications, partial SVD algorithms can be expensive, and a 
description in terms of actual columns and/or rows is of interest 
\cite{GTZ97,GT01}.
A final motivation for studying matrix decompositions of this form is to
obtain low-rank matrix approximations to extremely large matrices where a
computation of the SVD is too expensive 
\cite{FKV98,FKV04,dkm_matrix1,dkm_matrix2,dkm_matrix3}.

\subsection{Our Main Results}

Our main algorithmic results have to do with efficiently computing low-rank 
matrix approximations that are explicitly expressed in terms of a small 
number of columns and/or rows of the input matrix.
We start with the following definition.
\begin{definition}
\label{def:def_CX}
Let $A$ be an $m \times n$ matrix. For any given  
$C$, an $m \times c$ matrix whose 
columns consist of $c$ columns of the matrix $A$, 
the $m \times n$ matrix $A'= C X $ is a 
\emph{column-based matrix approximation} to $A$,
or \emph{CX matrix decomposition},
for any $c \times n$ matrix $X$. 
\end{definition}
Several things should be noted about this definition.
First, we will be interested in $c \ll n$ in our applications.
For example, depending on the application, $c$ could be constant, independent 
of $n$, logarithmic in the size of $n$, or simply a large constant factor less 
than $n$.
Second, a CX matrix decomposition expresses each of the columns of $A$ in terms 
of a linear combination of ``dictionary elements'' or ``basis columns,'' each 
of which is an actual column of $A$.
Thus, a CX matrix decomposition provides a low-rank approximation to the
original matrix, although one with structural properties that are quite
different than those provided by the SVD.
Third, given a set of columns $C$, the approximation $A' = P_CA = CC^+A$ 
(where $P_CA$ is the projection of $A$ onto the subspace spanned by the 
columns of $C$ and $C^+$ is the Moore-Penrose generalized inverse of $C$, 
as defined in Section~\ref{sxn:review_la})
clearly satisfies the requirements of Definition~\ref{def:def_CX}.
Indeed, this is the ``best'' such approximation to $A$, in the sense that 
$\FNorm{A - C\left(C^+A\right)}
   = \min_{X \in \mathbb{R}^{c \times n}} \FNorm{A - CX}$.

Our first main result is the following.
\begin{theorem}
\label{thm:thm_CX}
Given a matrix $A \in \mathbb{R}^{m \times n}$ and an integer
$k \ll \min \{m,n\}$, there exist randomized algorithms 
such that either exactly $c = O(k^2 \log (1/\delta)/\epsilon^2)$ columns of 
$A$ are chosen to construct $C$, or 
$c = O(k \log k \log (1/\delta)/\epsilon^2)$ columns 
are chosen in expectation to construct $C$,
such that with probability at least 
$1-\delta$, 
\begin{equation}
\label{eqn:eqn_prob_CX}
\min_{X \in \mathbb{R}^{c \times n}} \FNorm{A - CX} = 
\FNorm{A - CC^+A} 
   \leq (1+\epsilon) \FNorm{A-A_k}.
\end{equation}
Here, $C$ is a matrix consisting of the chosen columns of $A$, $CC^+A$ is the 
projection of $A$ on the subspace spanned by the chosen columns, and $A_k$ is 
the best rank-$k$ approximation to $A$. 
Both algorithms run in time $O(SVD(A,k))$, which is the 
time required to compute the best 
rank-$k$ approximation to the matrix $A$~\cite{GVL96}.
\end{theorem}
Note that we use $c > k$ and 
have an $\epsilon$ error, which allows us to take advantage of linear 
algebraic structure in order to obtain an efficient algorithm.
In general, this would not be the case if, given an $m \times n$ matrix $A$, 
we had specified a parameter $k$ and asked for the ``best'' subset of $k$ 
columns, where ``best'' is measured, e.g., by maximizing the Frobenius norm 
captured by projecting onto those columns or by maximizing the volume of the 
parallelepiped defined by those columns.
Also, it is not clear {\em a priori} that $C$ with properties above
even exists; see the discussion in 
Sections~\ref{sxn:previous:algs} and \ref{sxn:previous:relerr}.
Finally, 
our result does not include any reference to 
regularization or conditioning, as is common in certain application domains; 
a discussion of similar work on related problems in numerical linear algebra 
may be found in Section~\ref{sxn:previous:nla}.

Our second main result extends the previous result to CUR matrix decompositions. 


\begin{definition}
\label{def:def_CUR}
Let $A$ be an $m \times n$ matrix. 
For any given
$C$, an $m \times c$ matrix whose
columns consist of $c$ columns of the matrix $A$,
and $R$, an 
$r \times n$ matrix whose rows consist of $r$ rows of the 
matrix $A$,
the $m \times n$ matrix $A' = CUR $ is a 
\emph{column-row-based matrix approximation} to $A$, 
or \emph{CUR matrix decomposition},
for any $c \times r$ matrix $U$.
\end{definition}
Several things should be noted about this definition.
First, a CUR matrix decomposition is a CX matrix decomposition, but one with 
a very special structure, i.e., every column of $A$ can be expressed in terms 
of the basis provided by $C$ using only the information contained in a small 
number of rows of $A$ and a low-dimensional encoding matrix.
Second, in terms of its singular value structure, $U$ must clearly contain
``inverse-of-$A$'' information.
For the CUR decomposition described in this paper, $U$ will be a generalized 
inverse of the intersection between $C$ and $R$. 
More precisely, if $C=AS_CD_C$ and $R=D_RS_R^TA$ then
$U = (D_RS_R^TAS_CD_C)^+$.
(See Section~\ref{sxn:review_la} for a review of linear algebra and 
notation, such as that for $S_C$, $D_C$, $S_R$, and $D_R$.)
Third, the combined size of $C$, $U$ and $R$ is $O(mc+rn+cr)$, which is an 
improvement over $A$'s size of $O(mn)$ when $c,r \ll n,m$.
Finally, note the structural simplicity of a CUR matrix decomposition:
\begin{equation}
\underbrace{\left(
   \begin{array}{ccccc}
   &&&& \\
   &&&& \\
   &&A&& \\
   &&&& \\
   &&&&
   \end{array}
\right)}_{m \times n}
\approx
\underbrace{\left(
   \begin{array}{ccc} && \\  &&\\ &C& \\ && \\ && \end{array}
\right)}_{m \times c}
\underbrace{ \left(
   \begin{array}{ccc}  && \\ &U& \\ &&  \end{array}
\right) }_{c \times r}
\underbrace{\left(
   \begin{array}{ccccc}  &&&& \\ &&R&& \\ &&&&  \end{array}
\right)}_{r \times n} .
\end{equation}

Our main result for CUR matrix decomposition is the following. 

\begin{theorem}
\label{thm:thm_CUR}
Given a matrix $A \in \mathbb{R}^{m \times n}$ and an integer
$k \ll \min \{m,n\}$, there exist randomized algorithms 
such that exactly $c = O(k^2 \log (1/\delta)/\epsilon^2)$ columns of 
$A$ are chosen to construct $C$, and then exactly 
$r = O(c^2 \log (1/\delta)/\epsilon^2)$ rows of $A$ are chosen to construct 
$R$, or 
$c = O(k \log k \log (1/\delta)/\epsilon^2)$ columns 
of $A$ in expectation are chosen to construct $C$, and then
$r = O(c \log c \log (1/\delta)/\epsilon^2)$ rows of $A$ in expectation are 
chosen to construct $R$, such that
such that with probability at least 
$1-\delta$, 
\begin{equation}
\label{eqn:eqn_prob_CUR}
\FNorm{A - CUR} 
   \leq (1+\epsilon) \FNorm{A-A_k}.
\end{equation}
Here, the matrix $U$ is a weighted Moore-Penrose inverse of the
intersection between $C$ and $R$,
and $A_k$ is the best rank-$k$ approximation to $A$. 
Both algorithms run in time $O(SVD(A,k))$, which is the 
time required to compute the best 
rank-$k$ approximation to the matrix $A$~\cite{GVL96}.
\end{theorem}

\subsection{Summary of Main Technical Result}
\label{sxn:intro:technical_l2}

The key technical insight that leads to the relative-error guarantees is that
the columns are selected by a novel sampling procedure that we call
``subspace sampling.''
Rather than sample columns from $A$ with a probability distribution that
depends on the Euclidean norms of the columns of $A$ (which gives provable
additive-error bounds~\cite{dkm_matrix1,dkm_matrix2,dkm_matrix3}),
in ``subspace sampling'' we randomly sample columns of $A$ with a probability
distribution that depends on the Euclidean norms of the rows of the top $k$
right singular vectors of $A$.
This allows us to capture entirely a certain subspace of interest.
Let $V_{A,k}$ be the $n \times k$ matrix whose columns consist of the top $k$
right singular vectors of $A$.
The ``subspace sampling'' probabilities $p_i, i\in [n]$ will satisfy
\begin{equation}
\label{eqn:rel}
p_i \geq \frac{\beta \VTTNormS{\left(V_{A,k}\right)_{(i)}}}{k}
\qquad
\forall i \in [n] ,
\end{equation}
for some $\beta \in (0,1]$, where $\left(V_{A,k}\right)_{(i)}$ is the $i$-th
\emph{row} of $V_{A,k}$.
That is, we will sample based on the norms of the rows (not the columns) of 
the truncated matrix of singular vectors.
Note that $\sum_{j=1}^n \VTTNormS{\left(V_{A,k}\right)_{(j)}}=k$ and 
that $\sum_{i \in [n]} p_i = 1 $.
To construct sampling probabilities satisfying Condition~(\ref{eqn:rel}), it
is sufficient to spend $O(SVD(A,k))$ time to compute (exactly or
approximately, in which case $\beta=1$ or $\beta < 1$, respectively) the top
$k$ right singular vectors of $A$.
Sampling probabilities of this form will allow us to deconvolute subspace
information and ``size-of-$A$'' information in the input matrix $A$, which in
turn will allow us to obtain the relative-error guarantees we desire.
Note that we have used this method previously~\cite{DMM06}, but in that case
the sampling probabilities contained other terms that complicated their 
interpretation.

We will use these ``subspace sampling'' probabilities in our main technical
result, which is a random sampling algorithm for approximating the following
generalized version of the standard $\ell_2$ regression problem.
Our main column/row-based approximation algorithmic results will follow from 
this result.
Given as input a matrix $A \in \mathbb{R}^{m \times n}$ that has rank no more 
than $k$ and a matrix of target vectors $B \in \mathbb{R}^{m \times p}$, 
compute
\begin{equation}
\label{eqn:original_prob_gen}
{\cal Z} = \min_{X \in \mathbb{R}^{n \times p}} \FNorm{B - AX}.
\end{equation}
That is, fit every column of the matrix $B$ to the basis provided by the 
columns of the rank-$k$ matrix $A$.
Also of interest is the computation of 
\begin{equation}
\label{eqn:xopt_gen}
X_{opt} = A^+B.
\end{equation}
The main technical result of this paper is a simple sampling algorithm that 
represents the matrices $A$ and $B$ by a small number of rows so that this 
generalized $\ell_2$ regression problem can be solved to accuracy 
$1 \pm \epsilon$ for any $\epsilon > 0$.

More precisely, we present and analyze an algorithm 
(Algorithm~\ref{alg:sample_l2regression} of 
Section~\ref{sxn:generalized_l2_regression}) that constructs and solves 
an induced subproblem of the generalized $\ell_2$ regression problem of 
Equations (\ref{eqn:original_prob_gen}) and (\ref{eqn:xopt_gen}).
Let $DS^TA$ be the $r \times n$ matrix consisting of the sampled and 
appropriately rescaled rows of the original matrix $A$, and let $DS^TB$ be the 
$r \times p$ matrix consisting of the sampled and appropriately rescaled rows 
of $B$.
Then consider the problem
\begin{equation}
\label{eqn:sample_problem_gen}
\tilde{\cal Z}
   = \min_{X \in \mathbb{R}^{n \times p}} \FNorm{DS^TB - DS^TAX}     .
\end{equation}
The ``smallest'' matrix $\tilde{X}_{opt} \in \mathbb{R}^{n \times p}$ among 
those that achieve the minimum value $\tilde{\cal Z}$ in this sampled $\ell_2$
regression problem is
\begin{equation}
\label{eqn:xs_gen}
\tilde{X}_{opt} = \left(DS^TA\right)^+ DS^TB     .
\end{equation}
Since we will sample a number of rows $r \ll m$ of the original problem, we
will compute (\ref{eqn:xs_gen}), and thus (\ref{eqn:sample_problem_gen}), 
exactly.
Our main theorem, Theorem \ref{thm:ls_bound}, states that under appropriate
assumptions on the original problem and on the sampling probabilities, the
computed quantities $\tilde{\cal Z}$ and $\tilde{X}_{opt}$ will provide 
very accurate relative-error approximations to the exact solution 
${\cal Z}$ and the optimal vector $X_{opt}$.
Rows will be sampled with one of two random sampling procedures.
In one case, exactly $r = O(k^2/\epsilon^2)$ rows are chosen, and in the 
other case, $r = O(k \log k/\epsilon^2)$ rows in expectation are chosen.
In either case, the most expensive part of the computation involves the 
computation of the Euclidean norms of the rows of the right singular vectors
of $A$ which are used in the sampling probabilities.

\subsection{Outline of the Remainder of the Paper}
\label{sxn:intro:outline}

In the next two sections, we provide a review of relevant linear algebra, 
and we discuss related work.
Then, in Sections~\ref{sxn:mainCXresult} and~\ref{sxn:mainCURresult}, we 
present in detail our main algorithmic results.
In Section~\ref{sxn:mainCXresult}, we describe our main column-based matrix 
approximation algorithm, and 
in Section~\ref{sxn:mainCURresult}, we describe our main column-row-based 
matrix approximation algorithm.
Then, in Section~\ref{sxn:generalized_l2_regression}, we present an 
approximation algorithm for generalized $\ell_2$ regression.
This is our main technical result, and from it our two main algorithmic 
results will follow.
Finally, in Section~\ref{sxn:empirical} we present an empirical 
evaluation of our algorithms, and
in Section~\ref{sxn:conclusion} we present a brief conclusion.
We devote Appendix~\ref{sxn:matrix_multiply} to two prior algorithms for
approximate matrix multiplication.
These two algorithms select columns and rows in a complementary manner, and 
they are essential in the proof of our main results.

\section{Review of Linear Algebra}
\label{sxn:review_la}

In this section, we provide a review of linear algebra that will be useful 
throughout the paper; for more details, 
see~\cite{Nashed76,HJ85,Stewart90,GVL96,Bhatia97,BIG03}.
We also review a sampling matrix formalism that will be
convenient in our discussion~\cite{dkm_matrix1}.

Let $[n]$ denote the set $\{1,2,\ldots,n\}$.
For any matrix $A \in \mathbb{R}^{m \times n}$, let $A_{(i)}, i \in [m]$ 
denote the $i$-th row of $A$ as a row vector, and let $A^{(j)}, j \in [n]$
denote the $j$-th column of $A$ as a column vector.
In addition, let $ \FNormS{A} = \sum_{i=1}^m \sum_{j=1}^n A_{ij}^2 $ denote 
the square of its Frobenius norm, and let
$\TNorm{A} = \sup_{x\in \mathbb{R}^n,\ x\neq 0} \VTTNorm{Ax}/\VTTNorm{x}$
denote its spectral norm.
These norms satisfy: 
$\TNorm{A} \leq \FNorm{A} \leq \sqrt{\min\{m,n\}} \TNorm{A}$ for any matrix 
$A$; and also $\FNorm{AB} \leq \FNorm{A}\TNorm{B}$ for any matrices $A$ and 
$B$.

If $A \in \mathbb{R}^{m \times n}$, then there exist orthogonal matrices
$ U=[u^{1} u^{2} \ldots u^{m}]\in \mathbb{R}^{m \times m} $ and 
$ V=[v^{1} v^{2} \ldots v^{n}]\in \mathbb{R}^{n \times n} $
such that
$
U^TAV=\Sigma=\diag(\sigma_1,\ldots,\sigma_\xi)    ,
$
where $\Sigma \in \mathbb{R}^{m \times n}$, $\xi=\min\{m,n\}$ and 
$\sigma_1 \geq \sigma_2 \geq \ldots \geq \sigma_\xi \geq 0$.  
Equivalently, 
$ A =U{\Sigma}V^T $.  
The three matrices $U$, $V$, and $\Sigma$ constitute the Singular Value 
Decomposition (SVD) of $A$.
The $\sigma_i$ are the singular values of $A$, the vectors $u^{i}$, $v^{i}$ 
are the $i$-th left and the $i$-th right singular vectors of $A$, 
respectively, and the condition number of $A$ is 
$\kappa(A) = \sigma_{\max}(A)/\sigma_{\min}(A)$.
If $k \leq r = \mbox{rank}(A)$, then the SVD of $A$ may be written as
\begin{equation} 
\label{eqn:svdA}
A = U_A \Sigma_A V_A^T 
  = \left[\begin{array}{cc} 
          U_{k} & U_{k}^{\perp}
    \end{array}\right]
    \left[\begin{array}{cc}
          \Sigma_{k} & \bf{0}        \\
          \bf{0} & \Sigma_{k,\perp}
    \end{array}\right]
    \left[\begin{array}{c}
           V_{k}^T \\
           {V_{k}^{\perp}}^T
    \end{array}\right]     
  = U_k \Sigma_k V_k^T + U_{k}^{\perp} \Sigma_{k,\perp} {V_{k}^{\perp}}^T     .
\end{equation}
Here, $\Sigma_k$ is the $k \times k$ diagonal matrix containing the top $k$ 
singular values of $A$, and
$\Sigma_{k,\perp}$ is the $\left(r-k\right) \times \left(r - k\right)$ 
diagonal matrix containing the bottom $r-k$ nonzero singular values of $A$. 
Also, $V_{k}^T$ is the $k \times n$ matrix whose rows are the top $k$ right 
singular vectors of $A$, 
${V_{k}^{\perp}}^T$ is the $\left(r - k\right) \times n$ matrix whose rows are the 
bottom $r - k$ right singular vectors of $A$,
and $U_k$ and $U_{k}^{\perp}$ are defined similarly.
If we define $ A_k = U_k\Sigma_k V_k^T $, then the distance (as measured by 
both $\TNorm{\cdot}$ and $\FNorm{\cdot}$) between $A$ and any rank $k$ 
approximation to $A$ is minimized by $A_k$.
We will denote by $O(SVD(A,k))$ the time required to compute the best 
rank-$k$ approximation to the matrix $A$~\cite{GVL96}.
Finally, for any orthogonal matrix $U \in \mathbb{R}^{m \times c}$, let 
$U^{\perp} \in \mathbb{R}^{m \times (m-c)}$ denote an orthogonal matrix whose 
columns are an orthonormal basis spanning the subspace of $\mathbb{R}^m$ that 
is orthogonal to the column space of $U$.

Given a matrix $A \in \mathbb{R}^{m \times n}$, the unweighted Moore-Penrose 
generalized inverse of $A$, denoted by $A^{+}$, is the unique $n \times m$ 
matrix that satisfies the four Moore-Penrose conditions~\cite{Nashed76,BIG03}.
In terms of the SVD this generalized inverse may be written as
$A^+ = V_{A} \Sigma_{A}^{-1} U_{A}^T$ (where the square diagonal 
$\mbox{rank}(A) \times \mbox{rank}(A)$ matrix $\Sigma_{A}$, as in 
(\ref{eqn:svdA}), is invertible by construction).
If, in addition, $D_1 \in \mathbb{R}^{m \times m}$ and 
$D_2 \in \mathbb{R}^{n \times n}$ are diagonal matrices with positive entries 
along the diagonal, then the $\{D_1,D_2\}$-Moore-Penrose generalized inverse 
of $A$, denoted by $A^{+}_{(D_1,D_2)}$, is a generalization of the 
Moore-Penrose inverse that can be expressed in terms of the unweighted 
generalized inverse of $A$ as
$A^{+}_{(D_1,D_2)} = D_2^{-1/2} ( D_1^{1/2} A D_2^{-1/2} )^{+} D_1^{1/2} $.
Also, in terms of the generalized inverse, the projection onto the column 
space of any matrix $A$ may be written as $P_A = AA^+$.

Since our main algorithms will involve sampling columns and/or rows from 
input matrices (using one of two related random sampling procedures described 
in Appendix~\ref{sxn:matrix_multiply}), we conclude this subsection with a 
brief review of a sampling matrix formalism that was introduced in 
\cite{dkm_matrix1} and with respect to which our sampling matrix operations
may be conveniently expressed.
First, assume that $c'$ ($=c$ \emph{exactly}) columns of $A$ are 
chosen in $c$ i.i.d. trials by randomly sampling according to a probability 
distribution $\{p_i\}_{i=1}^{n}$ with the \textsc{Exactly($c$)} algorithm 
(described in detail in Appendix~\ref{sxn:matrix_multiply}), and
assume that the $i_t$-th column of $A$ is chosen in the $t$-th (for 
$t=1,\ldots,c$) independent random trial.
Then, define the sampling matrix $S \in \mathbb{R}^{n \times c}$ to be the 
zero-one matrix where $S_{i_tt} = 1$ and $S_{ij} = 0$ otherwise, and define 
the diagonal rescaling matrix $D \in \mathbb{R}^{c \times c}$ to be the 
diagonal matrix with $D_{tt} = 1/\sqrt{cp_{i_t}}$, where $p_{i_t}$ is the 
probability of choosing the $i_t$-th column.
Alternatively, assume that $c'$ ($\le c$ \emph{in expectation}) columns of
$A$ are chosen with the \textsc{Expected($c$)} 
algorithm (also described in detail in Appendix~\ref{sxn:matrix_multiply}) by 
including the $i$-th column of $A$ in $C$ with probability 
$\tilde{p}_i = \min \{1,cp_i\}$.
Then, define the sampling matrix $S \in \mathbb{R}^{n \times n}$ to be the 
zero-one matrix where $S_{ii} = 1$ if the $i$-th column is chosen and 
$S_{ij} = 0$ otherwise, and define the rescaling matrix 
$D \in \mathbb{R}^{n \times c'}$ to be the matrix with 
$D_{ij} = 1/\sqrt{c\tilde{p}_{j}}$ if $i-1$ of the previous columns have 
been chosen and $D_{ij}=0$ otherwise.
Clearly, in both of these cases, $C = ASD$ is an $m \times c'$ matrix 
consisting of sampled and rescaled copies of the columns of $A$, and 
$R = (SD)^{T}A = DS^TA$ is a $c' \times n$ matrix consisting of sampled and 
rescaled copies of the rows of $A$.
In certain cases, we will subscript $S$ and $D$ with $C$ or $R$ (e.g., 
$C = AS_CD_C$ and $R=D_RS_R^TA$) to make explicit that the corresponding 
sampling and rescaling matrices are operating on the columns or rows, 
respectively, of $A$.

\section{Relationship with Previous Related Work}
\label{sxn:previous}

In this section, we discuss the relationship between our results and related 
work in numerical linear algebra and theoretical computer science.

\subsection{Related Work in Numerical Linear Algebra}
\label{sxn:previous:nla}

Within the numerical linear algebra community, several groups have studied 
matrix decompositions with similar structural, if not algorithmic, properties 
to the CX and CUR matrix decompositions we have defined.
Much of this work is related to the QR decomposition, originally used 
extensively in pivoted form by Golub~\cite{Golub65,BG65}.

Stewart and collaborators were interested in computing sparse low-rank 
approximations to large sparse term-document 
matrices~\cite{Ste99,Ste04_TR,BPS04_TR}.
He developed the quasi-Gram-Schmidt method.  
This method is a variant of the QR decomposition which, when given as input an 
$m \times n$ matrix $A$ and a rank parameter $k$, returns an $m \times k$ 
matrix $C$ consisting of $k$ columns of $A$ whose span approximates the column 
space of $A$ and also a nonsingular upper-triangular $k \times k$ matrix $T_C$ 
that orthogonalizes these columns (but it does not explicitly compute the 
nonsparse orthogonal matrix $Q_C = C T_C^{-1}$).
This provides a matrix decomposition of the form $A \approx CX$.
By applying this method to $A$ to obtain $C$ and to $A^T$ to obtain an 
$k \times n$ matrix $R$ consisting of $k$ rows of $A$, one can show that
$A \approx CUR$, where the matrix $U$ is computed to minimize $\FNormS{A-CUR}$.
Although provable approximation guarantees of the form we present were not 
provided, backward error analysis was performed and the method was shown to 
perform well empirically~\cite{Ste99,Ste04_TR,BPS04_TR}.

Goreinov, Tyrtyshnikov, and Zamarashkin \cite{GTZ97,GT01,Tyr04b} were interested in 
applications such as scattering, in which large coefficient matrices have
blocks that can be easily approximated by low-rank matrices.
They show that if the matrix $A$ is approximated by a rank-$k$ matrix to 
within an accuracy $\epsilon$ then \emph{there exists} a choice of $k$ columns 
and $k$ rows, i.e., $C$ and $R$, and a low-dimensional $k \times k$ matrix $U$ 
constructed from the elements of $C$ and $R$, such that $A \approx CUR$ in the 
sense that $ \TNorm{A-CUR} \le \epsilon f(m,n,k)$, where 
$ f(m,n,k) = 1+ 2\sqrt{km} + 2\sqrt{kn} $.
In \cite{GTZ97}, the choice for these matrices is related to the problem of 
determining the minimum singular value $\sigma_k$ of $k \times k$ submatrices 
of $n \times k$ orthogonal matrices.
In addition: in \cite{GT01} the choice for $C$ and $R$ is interpreted in terms 
of the maximum volume concept from interpolation theory, in the sense that 
columns and rows should be chosen such that their intersection $W$ defines a 
parallelepiped of maximum volume among all $k \times k$ submatrices of $A$; 
and in \cite{Tyr04b} an empirically effective deterministic algorithm is 
presented which ensures that $U$ is well-conditioned.

Gu and Eisenstat, in their seminal paper \cite{GE96}, describe a strong 
rank-revealing QR factorization that deterministically selects exactly $k$ 
columns from an $m \times n$ matrix $A$.
The algorithms of \cite{GE96} are efficient, in that their running time is
$O(mn^2)$ (assuming that $m \geq n$), which is essentially the time required 
to compute the SVD of $A$.
In addition, Gu and Eisenstat prove that if the $m \times k$ matrix $C$ 
contains the $k$ selected columns (without any rescaling), then
$ \sigma_{\min}(C) \geq \sigma_k(A)/f(k,n) $, where $f(k,n)=O(\sqrt{k(n-k)})$.
Thus, the columns of $C$ span a parallelepiped whose volume (equivalently, the 
product of the singular values of $C$) is ``large.''
Currently, we do not know how to convert this property into a statement 
similar to that of Theorem~\ref{thm:thm_CX}, although perhaps this can be 
accomplished by relaxing the number of columns selected by the algorithms 
of~\cite{GE96} to $O(poly(k,1/\epsilon))$.
For related work prior to Gu and Eisenstat, see Chan and Hansen~\cite{CH90,CH92}.

Finally, very recently, Martinsson, Rokhlin, and Tygert~\cite{MRT06_TR2}
proposed another related method to efficiently compute an approximation to the
best rank-$k$ approximation of an $m \times n$ matrix $A$.
The heart of their algorithm is a random projection method, which projects $A$
to a small number -- say $\ell$ -- of random vectors; the entries of these
random vectors are i.i.d.  Gaussians of zero mean and unit variance.
The general form of their bounds is quite complicated, but by setting, e.g.,
$\ell = k+20$, they construct a rank-$k$ approximation $A'$ to $A$ such that
\begin{equation}
\label{eqn:interpolative}
\TNorm{A - A'} \leq 10\sqrt{\left(k+20\right)m} \TNorm{A - A_k}
\end{equation}
holds with probability at least $1-10^{-17}$.
In addition, the authors extend their algorithm to compute the so-called
interpolative decomposition of a matrix $A$.
This decomposition is explicitly expressed in terms of a small number of columns
of $A$, and is a more restrictive version of our CX matrix decomposition.
More specifically, it additionally requires that every entry of $X$ is bounded
in absolute value by a small constant (e.g., two).
Thus, their algorithm computes an interpolative approximation $A' = CX$ to
$A$, where $C$ has only $\ell = k+20$ columns -- as opposed to the
$O(k \log k)$ columns that are necessary in our work -- and satisfies the
bound of (\ref{eqn:interpolative}).
Notice that their work provides bounds for the spectral norm, whereas our work
focuses only on the Frobenius norm.
However, their bounds are much weaker than our relative error bounds, since
$\sqrt{m\left(k+20\right)} \TNorm{A-A_k}$ might in general be larger even than
$\FNorm{A}$.

\subsection{Related Work in Theoretical Computer Science}
\label{sxn:previous:algs}

Within the theory of algorithms community, much research has followed the 
seminal work of Frieze, Kannan, and Vempala~\cite{FKV98,FKV04}.
Their work may be viewed, in our parlance, as sampling columns from a matrix
$A$ to form a matrix $C$ such that
$\FNorm{A-CX}\leq \FNorm{A-A_k}+\epsilon \FNorm{A}$.
The matrix $C$ has $poly(k,1/\epsilon, 1/\delta)$ columns and is constructed
after making only two passes over $A$ using $O(m+n)$ work space.
Under similar resource constraints, a series of papers have
followed~\cite{FKV98,FKV04} in the past seven
years~\cite{DFKVV99,dkm_matrix2,RV03},
improving the dependency of $c$ on $k, 1/\epsilon$, and $1/\delta$, and
analyzing the spectral as well as the Frobenius norm, yielding bounds of the 
form
\begin{equation}
\label{eqn:bound}
\XNorm{A-CX}\leq \XNorm{A-A_k}+\epsilon \FNorm{A}
\end{equation}
for $\xi = 2,F$, and thus providing additive-error guarantees for column-based
low-rank matrix approximations.

Additive-error approximation algorithms for CUR matrix decompositions have 
also been analyzed by Drineas, Kannan, and 
Mahoney~\cite{DK03,dkm_matrix1,dkm_matrix2,dkm_matrix3,dm_kernel_CONF,dm_kernel_JRNL}. 
In particular, in \cite{dkm_matrix3}, they compute an approximation to an 
$m \times n$ matrix $A$ by sampling $c$ columns and $r$ rows from $A$ to form 
$m \times c$ and $r \times n$ matrices $C$ and $R$, respectively.
From $C$ and $R$, a $c \times r$ matrix $U$ is constructed such that under 
appropriate assumptions
\begin{equation}
\XNorm{A-CUR} \le \XNorm{A-A_k} + \epsilon \FNorm{A}     ,
\label{eqn:cur_decomp_bound}
\end{equation}
with high probability, for both the spectral and Frobenius norms, $\xi=2,F$.
In \cite{dm_kernel_CONF,dm_kernel_JRNL}, it is further shown that if $A$ is a 
symmetric positive semidefinite (SPSD) matrix, then one can choose $R=C^T$ and 
$U=W^{+}$, where $W$ is the $c \times c$ intersection between $C$ and $R=C^T$,
thus obtaining an approximation $A \approx A' = CW^{+}C^T$.
This approximation is SPSD and has provable bounds of the form
(\ref{eqn:cur_decomp_bound}), except that the scale of the additional 
additive error is somewhat larger \cite{dm_kernel_CONF,dm_kernel_JRNL}.

Most relevant for our relative-error CX and CUR matrix decomposition 
algorithms is the recent work of Rademacher, Vempala and Wang~\cite{RVW05} and
Deshpande, Rademacher, Vempala and Wang~\cite{DRVW06}.
Using two different methods (in one case iterative sampling in a backwards
manner and an induction on $k$ argument~\cite{RVW05} and in the other case an
argument which relies on 
estimating the volume of the simplex formed by each of the $k$-sized subsets 
of the columns~\cite{DRVW06}), they reported the \emph{existence} of a set of 
$O(k^2/\epsilon^2)$ columns that provide relative-error CX matrix
decomposition.
No algorithmic result was presented, except for an exhaustive algorithm that
ran in $\Omega(n^k)$ time.
Note that their results did not apply to columns and rows \emph{simultaneously}.
Thus, ours is the first CUR matrix decomposition algorithm with relative 
error, and it was previously not even known whether such a relative-error 
$CUR$ representation existed, i.e., it was not previously known whether 
columns and rows satisfying the conditions of Theorem~\ref{thm:thm_CUR}
existed.

Other related work includes that of Rudelson and 
Vershynin~\cite{Rud99,V03,RV06_DRAFT}, who provide an algorithm for CX matrix 
decomposition which has an improved additive error spectral norm bound of 
the form
$$
\TNorm{A-CX}\leq \TNorm{A-A_k}+\epsilon \sqrt{\TNorm{A}\FNorm{A}}    .
$$
Their proof uses an elegant result on random vectors in the isotropic
position~\cite{Rud99}, and since we use a variant of their result, it is 
described in more detail in Appendix~\ref{sxn:matrix_multiply}.
Achlioptas and McSherry have computed low-rank matrix approximations using 
sampling techniques that involve zeroing-out and/or quantizing individual 
elements \cite{AM01,AM03}.
The primary focus of their work was in introducing methods to accelerate 
orthogonal iteration and Lanczos iteration methods, and their analysis relied 
heavily on ideas from random matrix theory \cite{AM01,AM03}.
Agarwal, Har-Peled, and Varadarajan have analyzed so-called ``core sets'' as a 
tool for efficiently approximating various extent measures of a point 
set~\cite{AHV04,AHV06}.
The choice of columns and/or rows we present are a ``core set'' for 
approximate matrix computations; in fact, our algorithmic solution to 
Theorem~\ref{thm:thm_CX} solves an open question in their 
survey~\cite{AHV06}.
The choice of columns and rows we present may also be viewed as a set of 
variables and features chosen from a data matrix~\cite{BL97,CGKRS00,GE03}.
``Feature selection'' is a broad area that addresses the choice of columns 
explicitly for dimension reduction, but the metrics there are typically 
optimization-based~\cite{CGKRS00} or machine-learning based \cite{BL97}.
These formulations tend to have set-cover like solutions and are incomparable 
with the linear-algebraic structure such as the low-rank criteria we consider 
here that is common among data analysts.

\subsection{Very Recent Work on Relative-Error Approximation Algorithms}
\label{sxn:previous:relerr}

To the best of our knowledge, the first nontrivial \emph{algorithmic result} 
for relative-error low-rank matrix approximation was provided by a 
preliminary version of this paper \cite{DMM06_relerr_110305,DMM06_relerr_TR}.
In particular, an earlier version of Theorem~\ref{thm:thm_CX} provided the 
first known relative-error column-based low-rank approximation in polynomial 
time~\cite{DMM06_relerr_110305,DMM06_relerr_TR}. 
The major difference between our Theorem~\ref{thm:thm_CX} and our result 
in~\cite{DMM06_relerr_110305,DMM06_relerr_TR} is that the sampling 
probabilities in~\cite{DMM06_relerr_110305,DMM06_relerr_TR} are more 
complicated.
(See Section~\ref{sxn:main_l2_alg:discussion} for details on this.)
The algorithm from \cite{DMM06_relerr_110305,DMM06_relerr_TR} runs in 
$O(SVD(A,k))$ time (although it was originally reported to run in only 
$O(SVD(A))$ time), and it has a sampling complexity of 
$O(k^2\log(1/\delta)/\epsilon^2)$ columns.

Subsequent to the completion of the preliminary version of this 
paper~\cite{DMM06_relerr_110305,DMM06_relerr_TR}, several developments
have been made on relative-error low-rank matrix approximation algorithms.
First, Har-Peled reported an algorithm that takes as input an $m \times n$ 
matrix $A$, and in roughly $O(mnk^2 \log k)$ time returns as output a rank-$k$ 
matrix $A'$ with a relative-error approximation 
guarantee~\cite{HarPeled06_relerr_DRAFT}. 
His algorithm uses geometric ideas and involves sampling and merging 
approximately optimal $k$-flats; it is not clear if this approximation can be 
expressed in terms of a small number of columns of $A$.
Then, Deshpande and Vempala \cite{DV06_relerr_TR} reported an algorithm that 
takes as input an $m \times n$ matrix $A$ that also returns a relative-error 
approximation guarantee.  
Their algorithm extends ideas from \cite{RVW05,DRVW06}, and it leads to a 
CX matrix decomposition consisting of $O(k \log k)$ columns of $A$.
The complexity of their algorithm is $O(Mk^2 \log k)$, where $M$ is the 
number of nonzero elements of $A$, and their algorithm can be implemented in 
a data streaming framework with $O(k \log k)$ passes over the data.
In light of these developments, we simplified and generalized our preliminary 
results~\cite{DMM06_relerr_110305,DMM06_relerr_TR}, and we performed a more 
refined analysis to improve our sampling complexity to 
$O(k \log k )$. 
Most recently, we learned of work by Sarlos~\cite{Sarlos06}, who used
ideas from the recently developed fast Johnson-Lindenstrauss transform of
Ailon and Chazelle~\cite{AC06} to yield further improvements to a CX matrix
decomposition.

\section{Our Main Column-Based Matrix Approximation Algorithm}
\label{sxn:mainCXresult}

In this section, we describe an algorithm and a theorem, from which our first
main result, Theorem~\ref{thm:thm_CX}, will follow.

\subsection{Description of the Algorithm}
\label{sxn:mainCXresult:alg}

Algorithm~\ref{alg:algCX} takes as input an $m \times n$ matrix $A$, a rank 
parameter $k$, and an error parameter $\epsilon$.
It returns as output an $m \times c$ matrix $C$ consisting of a small number 
of columns of $A$.
The algorithm is very simple: sample a small number of columns according to 
a carefully-constructed nonuniform probability distribution.
Algorithm~\ref{alg:algCX} uses the sampling probabilities 
\begin{equation}
\label{eqn:colprob_cols_exact} 
p_i = \frac{1}{k} \VTTNormS{\left(V_{A,k}^T\right)^{(i)}}       ,
   \hspace{0.25in}
   \mbox{   }
   \forall i \in [n]   ,
\end{equation}
but it will be clear from the analysis of 
Section~\ref{sxn:generalized_l2_regression} that any sampling probabilities
such that $p_i \geq \beta\VTTNormS{\left(V_{A,k}^T\right)^{(i)}}/k$,
for some $\beta \in (0,1]$, will also work with a small $\beta$-dependent 
loss in accuracy.
Note that Algorithm~\ref{alg:algCX} actually consists of two related 
algorithms, depending on how exactly the columns are chosen.
The \textsc{Exactly($c$)} algorithm picks exactly $c$ columns of $A$ to be 
included in $C$ in $c$ i.i.d. trials, where in each trial the $i$-th column 
of $A$ is picked with probability $p_i$. 
The \textsc{Expected($c$)} algorithm picks in expectation at most $c$ 
columns of $A$ to create $C$, by including the $i$-th column of $A$ in $C$ 
with probability $\min\left\{1, cp_i\right\}$. 
See Algorithms \ref{alg:SDconstruct_exact} and \ref{alg:SDconstruct_expected} 
in Appendix~\ref{sxn:matrix_multiply} for more details about these two 
column-sampling procedures.

\begin{algorithm}[h]

\begin{framed}

\SetLine

\AlgData{
$A \in \mathbb{R}^{m \times n}$, 
a rank parameter $k$, and 
an error parameter $\epsilon$.
}

\AlgResult{
$C \in \mathbb{R}^{m \times c}$
}

\begin{itemize}
\item
Compute sampling probabilities $p_i$ for all $i \in [n]$ given by
(\ref{eqn:colprob_cols_exact})\;
\item
(Implicitly) construct a sampling matrix $S_C$ and a diagonal rescaling 
matrix $D_C$ with the \textsc{Exactly($c$)} algorithm or with the 
\textsc{Expected($c$)} algorithm\;
\item
Construct and return the matrix $C=AS_CD_C$ consisting of a small number of 
rescaled columns of $A$\;

\end{itemize}

\end{framed}

\caption{
A randomized algorithm for CX matrix decomposition.
}

\label{alg:algCX}

\end{algorithm}

The running time of Algorithm~\ref{alg:algCX} is dominated by the computation 
of the sampling probabilities (\ref{eqn:colprob_cols_exact}), for which 
$O(SVD(A,k))$ time suffices.
The top $k$ right singular vectors of $A$ can be efficiently (approximately) 
computed using standard algorithms~\cite{GVL96,Parlett98}.
The building block of these algorithms is a series of matrix-vector
multiplications, where the input matrix $A$ is iteratively multiplied with a
changing set of $k$ orthogonal vectors.
In each iteration (which can be implemented by making passes over the input 
matrix $A$), the accuracy of the approximation improves.
Even though the number of iterations required to bound the error depends on
quantities such as the gap between the singular values of $A$, these
algorithms work extremely well in practice.
As such, they are often treated as ``black boxes'' for SVD computation in the 
theoretical computer science literature; see, e.g.,~\cite{AM01, AM03}.

\subsection{Statement of the Theorem}
\label{sxn:mainCXresult:thm}

Theorem~\ref{thm:relerr_CX} is our main quality-of-approximation result for
Algorithm~\ref{alg:algCX}.

\begin{theorem}
\label{thm:relerr_CX}
Let $A \in \mathbb{R}^{m \times n}$, let $k$ be a rank parameter, and 
let $\epsilon \in (0,1]$.
If we set $c = 3200k^2/\epsilon^2$ and run Algorithm~\ref{alg:algCX} by 
choosing exactly $c$ columns from $A$ with the \textsc{Exactly($c$)} 
algorithm, then with probability at least $0.7$
\begin{equation}
\label{eqn:thm_relerr_CX}
\FNorm{A-CC^+A} \le (1+\epsilon) \FNorm{A-A_k}
\end{equation}
Similarly, if we set $c = O(k \log k/\epsilon^2)$ and run 
Algorithm~\ref{alg:algCX} by choosing no more than $c$ columns in 
expectation from $A$ with the \textsc{Expected($c$)} algorithm, then 
(\ref{eqn:thm_relerr_CX}) holds with probability at least $0.7$. 
\end{theorem}
\begin{Proof}
Since for every set of columns $C=AS_CD_C$, $X_{opt}=C^+A$ is the matrix that
minimizes $\FNorm{A-CX}$, it follows that 
\begin{eqnarray}
\FNorm{A-CC^+A} 
\nonumber
   &=&   \FNorm{A-(AS_CD_C)(AS_CD_C)^+A}     \\
\label{eqn1:prf:thm:relerr_CX}
   &\le& \FNorm{A-(AS_CD_C)(P_{A,k}AS_CD_C)^+P_{A,k}A}  ,
\end{eqnarray}
where $P_{A,k} = U_{A,k}U_{A,k}^T$ is a projection onto the top $k$ left 
singular vectors of $A$.
To bound (\ref{eqn1:prf:thm:relerr_CX}), consider the problem of approximating 
the solution to $ \min_{X \in \mathbb{R}^{m \times m}} \FNorm{XA_k-A}$ by 
randomly sampling columns of $A_k$ and of $A$.
It follows as a corollary of (\ref{eqn:result2}) of
Theorem~\ref{thm:ls_bound} of Section~\ref{sxn:generalized_l2_regression} that
\begin{equation}
\FNorm{A-(AS_CD_C)(A_kS_CD_C)^+A_k} 
   \le (1+\epsilon) \FNorm{A-AA_k^+A_k}   
   = (1+\epsilon) \FNorm{A-A_k}   ,
\label{eqn2:prf:thm:relerr_CX}
\end{equation}
which, when combined with (\ref{eqn1:prf:thm:relerr_CX}), establishes the 
theorem.
\end{Proof}

\textbf{Remark:}
For simplicity of presentation, we have presented Algorithm~\ref{alg:algCX} 
and Theorem~\ref{thm:relerr_CX} such that (\ref{eqn:thm_relerr_CX}) holds 
with only constant probability, but this can be boosted to hold with 
probability at least $1-\delta$ using standard methods.
In particular, consider the following:
run Algorithm~\ref{alg:algCX} (using either the \textsc{Exactly($c$)} algorithm
or the \textsc{Expected($c$)} algorithm, but with the appropriate value of $c$)
independently $\ln(1/\delta)$ times;
and return the $C$ such that $\FNorm{A-CC^+A}$ is smallest.
Then, since in each trial the claim of Theorem~\ref{thm:relerr_CX} fails with 
probability less than $0.3 < 1/e$, the claim of Theorem~\ref{thm:relerr_CX} 
will fail for every trial with probability less than 
$(1/e)^{\ln(1/\delta)} = \delta$.
This establishes Theorem~\ref{thm:thm_CX}.

\textbf{Remark:}
For simplicity of presentation, we have also 
stated Theorem~\ref{thm:relerr_CX} in such 
a way that the rank of the approximating matrix $A' = CC^+A$ may be greater 
than $k$.
This possibility may be undesirable in certain applications, and it can be 
easily removed.
Let $A^{''} = C(P_{A,k}C)^+P_{A,k}A$.
Then, it follows from (\ref{eqn2:prf:thm:relerr_CX}) that $A^{''}$ is a CX 
matrix approximation that is within relative error $\epsilon$ of the best 
rank-$k$ approximation to $A$ and that has rank no more than $k$.

\subsection{Discussion of the Analysis}
\label{sxn:mainCXresult:discuss}

Given a matrix $A$, Theorem~\ref{thm:thm_CX} asks us to find a set of 
columns $C = AS_CD_C$ such that $CC^+A$ ``captures'' almost as much of $A$ 
as does $A_k = U_{A,k}U_{A,k}^TA$.
Given that set (or any other set) of columns $C$, it is well-known that the 
matrix 
$X_{opt}=C^+A$ is the ``smallest'' matrix among those that solve the 
optimization problem (\ref{eqn:l2reg}).
For a given $A$ and $C$, let us approximate $X_{opt}$ as
$$
X_{opt} = C^+A \approx \left( P_{A,k} C \right)^+ P_{A,k} A  .
$$
This approximation is suboptimal with respect to solving the optimization 
problem (\ref{eqn:l2reg}), i.e., 
$$
\FNorm{A-CC^+A} 
   \le \FNorm{A-C\left(P_{A,k} C \right)^+ P_{A,k} A}   ,
$$
but it can be shown that by choosing $C$ properly, i.e., by choosing $S_C$ and 
$D_C$ (the column sampling and rescaling matrices) properly, we have that
$$
\FNorm{A-C\left( P_{A,k} C\right)^+ P_{A,k} A} 
   \le \left(1+\epsilon\right)\FNorm{A-A_k}   .
$$
The main technical challenge is to sample in a manner such that the 
column-sampled version of the matrix consisting of the top $k$ right singular 
vectors of $A$ is full rank, i.e., 
$\mbox{rank}(V_{A,k}^TS_CD_C) = \mbox{rank}(V_{A,k}^T) = k$.
To accomplish this, we sample with respect to probabilities of the form 
(\ref{eqn:colprob_cols_exact}).
To understand these sampling probabilities, recall that we seek to pick 
columns that span almost the same subspace as the top $k$ left singular 
vectors of $A$ (i.e., $U_k$), and recall that the $i$-th column of $A$ is 
equal to 
$$
A^{(i)} = U_k \Sigma_k \left(V_k^T\right)^{(i)} 
         + U_{\rho-k} \Sigma_{\rho-k} \left(V_{\rho-k}^T\right)^{(i)}  .
$$
Since post-multiplying $U_k$ by $\Sigma_k$ does not change the span of the 
columns of $U_k$, $\VTTNormS{\left(V_{k}^T\right)^{(i)}}$ measures ``how much'' 
of the $i$-th column of $A$ lies in the span of $U_{A,k}$, independent of 
the magnitude of the singular values associated with those directions.

\section{Our Main Column-Row-Based Matrix Approximation Algorithm}
\label{sxn:mainCURresult}

In this section, we describe an algorithm and a theorem that, when combined
with the results of Section~\ref{sxn:mainCXresult}, will establish our second
main result, Theorem~\ref{thm:thm_CUR}.

\subsection{Description of the Algorithm}
\label{sxn:mainCURresult:alg}

Algorithm~\ref{alg:algCUR} takes as input an $m \times n$ matrix $A$, an 
$m \times c$ matrix $C$ consisting of a small number of columns of $A$, and 
an error parameter $\epsilon$.
It returns as output an $r \times n$ matrix $R$ consisting of a small number 
of rows of $A$ and an $r \times c$ matrix $W$ consisting of the corresponding 
rows of $C$.
The algorithm is very simple: sample a small number of rows according to 
a carefully-constructed nonuniform probability distribution.
Algorithm~\ref{alg:algCUR} uses the sampling probabilities 
\begin{equation}
\label{eqn:colprob_rows_exact} 
p_i = \frac{1}{c} \VTTNormS{\left(U_{C}^T\right)^{(i)}}     ,
   \hspace{0.25in}
   \mbox{   }
   \forall i \in [m]   ,
\end{equation}
but it will be clear from the analysis of 
Section~\ref{sxn:generalized_l2_regression} that any sampling probabilities
$p_i, i \in [m]$, such that 
$p_i \geq \beta\VTTNormS{\left(U_{C}^T\right)^{(i)}}/c$,
for some $\beta \in (0,1]$, will also work with a small $\beta$-dependent 
loss in accuracy.
Note that Algorithm~\ref{alg:algCUR} actually consists of two related 
algorithms, depending on how exactly the rows are chosen.
The \textsc{Exactly($c$)} algorithm picks exactly $r$ rows of $A$ to be 
included in $R$ in $r$ i.i.d. trials, where in each trial the $i$-th row of 
$A$ is picked with probability $p_i$. 
The \textsc{Expected($c$)} algorithm picks in expectation at most $r$ rows of 
$A$ to create $R$, by including the $i$-th column of $A$ in $C$ with 
probability $\min\left\{1, rp_i\right\}$. 
See Algorithms \ref{alg:SDconstruct_exact} and \ref{alg:SDconstruct_expected} 
in Appendix~\ref{sxn:matrix_multiply} for more details about these two 
row-sampling procedures.

\begin{algorithm}[h]

\begin{framed}

\SetLine

\AlgData{
$A \in \mathbb{R}^{m \times n}$, 
$C \in \mathbb{R}^{m \times c}$ consisting of $c$ columns of $A$, 
a positive integer $r$, and
an error parameter $\epsilon$.
}

\AlgResult{
$R \in \mathbb{R}^{r \times n}$ consisting of $r$ rows of $A$ and
$W \in \mathbb{R}^{c \times r}$ consisting of the corresponding $r$ rows of $C$, and
$U \in \mathbb{R}^{r \times c}$.
}

\begin{itemize}
\item
Compute sampling probabilities $p_i$ for all $i \in [m]$ given by
(\ref{eqn:colprob_rows_exact})\; 
\item
(Implicitly) construct a sampling matrix $S_R$ and a diagonal rescaling 
matrix $D_R$ with the \textsc{Exactly($c$)} algorithm or with the 
\textsc{Expected($c$)} algorithm\;
\item
Construct and return the matrix $R=D_RS_R^TA$ consisting of a small number of 
rescaled rows of $A$\;
\item
Construct and return the matrix $W=D_RS_R^TC$ consisting of the corresponding 
rescaled rows of $C$\;
\item
Let $U=W^+$\;
\end{itemize}

\end{framed}

\caption{
A randomized algorithm for CUR matrix decomposition.
}

\label{alg:algCUR}

\end{algorithm}

Reading the input matrices to Algorithm~\ref{alg:algCUR} takes $O(mn)$ time;
computing the full SVD of $C$ requires $O(c^2m)$ time;
constructing the matrix $R$ requires $O(rn)$ time; 
constructing the matrix $W$ requires $O(rc)$ time; and 
computing $U$ requires $O(c^2r)$ time.
Overall, the running time of the algorithm is $O(mn)$ since $c, r$ are 
constants independent of $m, n$.
This can be improved if the input matrices are sparse, but for simplicity we 
omit this discussion.

\subsection{Statement of the Theorem}
\label{sxn:mainCURresult:thm}

Theorem~\ref{thm:relerr_CUR} is our main quality-of-approximation result for
Algorithm~\ref{alg:algCUR}.

\begin{theorem}
\label{thm:relerr_CUR}
Let $A \in \mathbb{R}^{m \times n}$, let $C \in \mathbb{R}^{m \times c}$ be a 
matrix consisting of any $c$ columns of $A$, and let $\epsilon \in (0,1]$.
If we set $r = 3200c^2/\epsilon^2$ and run Algorithm~\ref{alg:algCUR} by 
choosing $r$ rows exactly from $A$ and from $C$ with the \textsc{Exactly($c$)} 
algorithm, then with probability at least $0.7$
\begin{equation}
\label{eqn:thm_relerr_CUR}
\FNorm{A-CUR} \le (1+\epsilon) \FNorm{A-CC^+A}
\end{equation}
Similarly, if we set $r = O(c \log c/\epsilon^2)$ and run 
Algorithm~\ref{alg:algCUR} by choosing no more than $r$ rows in 
expectation from $A$ and from $C$ with the \textsc{Expected($c$)} algorithm, 
then (\ref{eqn:thm_relerr_CUR}) holds with probability at least $0.7$. 
\end{theorem}
\begin{Proof}
Consider the problem of approximating the solution to
$ \min_{X \in \mathbb{R}^{c \times n}} \FNorm{CX-A}$
by randomly sampling rows from $C$ and $A$.
It follows as a corollary of (\ref{eqn:result2}) of
Theorem~\ref{thm:ls_bound} of Section~\ref{sxn:generalized_l2_regression} that
$$
\FNorm{A - C (D_RS_R^TC)^+ D_RS_R^TA} \le (1+\epsilon) \FNorm{A-CC^+A}   ,
$$
where $R=D_RS_R^TA$ and $U=(D_RS_R^TC)^+$,
which establishes the theorem.
\end{Proof}

\textbf{Remark:}
For simplicity of presentation, we have presented Algorithm~\ref{alg:algCUR} 
and Theorem~\ref{thm:relerr_CUR} such that (\ref{eqn:thm_relerr_CUR}) holds 
with only constant probability, but this can be boosted to hold with 
probability at least $1-\delta$ using standard methods.
In addition, this can be combined with Algorithm~\ref{alg:algCX} and 
Theorem~\ref{thm:relerr_CX} by doing the following:
run Algorithm~\ref{alg:algCX} $\ln(2/\delta)$ times, and return the best $C$;
then, with that $C$ run Algorithm~\ref{alg:algCUR} $\ln(2/\delta)$ times, and 
return the best $U,R$ pair.
Then
$$
\FNorm{A-CUR} \le (1+\epsilon)  \FNorm{A-CC^+A}
              \le (1+\epsilon)^2\FNorm{A-A_k}
              \le (1+\epsilon') \FNorm{A-A_k}   ,
$$
where $\epsilon^{'} = 3 \epsilon$, and the combined failure probability is no 
more than $\delta/2+\delta/2 = \delta$.
This establishes Theorem~\ref{thm:thm_CUR}.

\subsection{Discussion of the Analysis}
\label{sxn:mainCURresult:discuss}

Assume that we are given an $m \times c$ matrix $C$, consisting of any set of 
$c$ columns of an $m \times n$ matrix $A$, and consider the following idea for 
approximating the matrix $A$.
The columns of $C$ are a set of ``basis vectors'' that are, in general, 
neither orthogonal nor normal. 
To express all the columns of $A$ as linear combinations of the columns of
$C$, we can solve
$$
\min_{x_j \in \mathbb{R}^c}\VTTNorm{A^{(j)} - Cx_j}     ,
$$
for each column $A^{(j)}, j \in [n]$, in order to find a $c$-vector of 
coefficients $x_j$ and get the optimal least-squares fit for $A^{(j)}$. 
Equivalently, we can solve an optimization problem of the form 
(\ref{eqn:l2reg}).
Note that if $m$ and $n$ are large and $c = O(1)$, then this is an 
overconstrained least-squares fit problem. 
It is well-known that $X_{opt} = C^+A$ is the ``smallest'' matrix solving 
this optimization problem, in which case we are using information from every 
row of $A$ to compute the optimal coefficient matrix.
Let us approximate $X_{opt}$ as
$$
X_{opt} = C^+A \approx \left(D_RS_R^TC\right)^+ D_RS_R^TA = \tilde{X}_{opt},
$$
and note that $\tilde{X}_{opt} = W^+ R$.
This matrix $\tilde{X}_{opt}$ is clearly suboptimal with respect to solving 
the optimization problem (\ref{eqn:l2reg}), i.e., 
$$
\FNorm{A-CC^+A} \le \FNorm{A-CW^+R}    ,
$$
but it can be shown that by choosing $S_R$ and $D_R$ (the row sampling and 
rescaling matrices) properly we have that
$$
\FNorm{A-CW^+R} \le \left(1+\epsilon\right)\FNorm{A-CC^+A}.
$$
As in Section~\ref{sxn:mainCXresult:discuss}, the main technical challenge 
is to sample in a manner such that the row-sampled version of the matrix 
consisting of the top $c$ left singular vectors of $C$ is full rank, i.e., 
$\mbox{rank}(D_RS_R^TU_{C,c}) = \mbox{rank}(U_{C,c}) = c$.

\section{An Approximation Algorithm for Generalized $\ell_2$ Regression}
\label{sxn:generalized_l2_regression}

The basic linear-algebraic problem of $\ell_2$ regression is one of the most 
fundamental regression problems, and it has found many applications in 
mathematics and statistical data analysis.
Recall the standard $\ell_2$ regression (or least-squares fit) problem:
given as input a matrix $A \in \mathbb{R}^{m \times n}$ and a target vector
$b \in \mathbb{R}^m$, compute
$
{\cal Z} = \min_{x \in \mathbb{R}^n} \VTTNorm{b - Ax}
$.
Also of interest is the computation of vectors that achieve the minimum
${\cal Z}$.
If $m > n$ there are more constraints than variables and the problem is an
overconstrained least-squares fit problem;
in this case, there does not in general exist a vector $x$ such that $Ax=b$.
It is well-known that the minimum-length vector among those minimizing
$\VTTNorm{b-Ax}$ is $ x_{opt} = A^+b $.
We previously presented an elaborate sampling algorithm that represents the 
matrix $A$ by a matrix by a small number of rows so that this $\ell_2$ 
regression problem can be solved to accuracy $1 \pm \epsilon$ for any 
$\epsilon > 0$ \cite{DMM06}.

This problem is of interest for CX and CUR matrix decomposition for the
following reason.
Given a matrix $A$ and a set of its columns $C$, if we want to get the best 
fit for every column of $A$ in terms of that basis, we want to solve 
$ CX \approx A $ for the matrix $X$.
More precisely, we would like to solve the optimization problem such as
\begin{equation}
\label{eqn:l2reg}
\mathcal{Z} = \min_{X \in \mathbb{R}^{c \times n}} \FNorm{A - CX}     .
\end{equation}
It is well-known that the matrix $X=C^+A$ is the ``smallest'' matrix among 
those that solve this problem.
In this case, we are approximating the matrix $A$  as $A' = CC^+A = P_C A$, 
and by keeping only the columns $C$ we are incurring an error of 
$\FNorm{A-CC^+A}$.
Two questions arise:
\begin{itemize}
\item
First, how do we choose the columns $C$ such that $\FNorm{A-CC^+A}$
is within relative error $\epsilon$ of  $\FNorm{A-A_k}$?
\item
Second, how do we choose the rows $R$ and a matrix $U$ such that 
$\FNorm{A-CUR}$ is within relative error $\epsilon$  of $\FNorm{A-CC^+A}$?
\end{itemize}
Motivated by these observations, we will consider the generalized version of
the standard $\ell_2$ regression problem, as defined in 
(\ref{eqn:original_prob_gen}) and (\ref{eqn:xopt_gen}).

In this section, we first present Algorithm~\ref{alg:sample_l2regression}, 
which is our main random sampling algorithm for approximating the solution to 
the generalized $\ell_2$ regression problem, and
Theorem~\ref{thm:ls_bound}, which provides our main quality-of-approximation 
bound for Algorithm~\ref{alg:sample_l2regression}.
Then, we discuss the novel nonuniform ``subspace sampling'' probabilities 
used by the algorithm.
Finally, we present the proof of Theorem~\ref{thm:ls_bound}.

\subsection{Description of the Algorithm and Theorem}
\label{sxn:main_l2_alg:alg_thm}

Algorithm~\ref{alg:sample_l2regression} takes as input an $m \times n$ matrix 
$A$ with rank no greater than $k$, an $m \times p$ matrix $B$, a set of 
sampling probabilities $\{p_i\}_{i=1}^m$, and a positive integer $r \le m$.
It returns as output a number $\tilde{\cal Z}$ and a $n \times p$ matrix
$\tilde{X}_{opt}$.
Using the sampling matrix formalism described in 
Section~\ref{sxn:review_la}, the algorithm (implicitly) forms a 
sampling matrix $S$, the transpose of which samples a few rows of $A$ and 
the corresponding rows of $B$, and a rescaling matrix $D$, which is a matrix 
scaling the sampled rows of $A$ and $B$. 
Since $r$ rows of $A$ and the corresponding $r$ rows of $B$ are sampled, 
the algorithm randomly samples $r$ of the $m$ constraints in the original 
$\ell_2$ regression problem.
Thus, the algorithm approximates the solution of the regression problem 
$AX \approx B$, as formalized in (\ref{eqn:l2reg}) and 
(\ref{eqn:original_prob_gen}), with the exact solution of the downsampled 
regression problem $DS^TA X \approx DS^TB$.
Note that it is the space of constraints that is sampled and 
that the dimensions of the unknown matrix $X$ are the same in both problems. 
Note also that although both $m$ and $n$ are permitted to be large, 
the problem is effectively overconstrained since $\mbox{rank}(A) \le k$.
As we will see below, $r=O(k \log k)$ or $r=O(k^2)$, depending on exactly how 
the random sample is constructed.
Thus, we will compute the solution to the sampled problem exactly.

\begin{algorithm}

\begin{framed}

\SetLine

\AlgData{
$A \in \mathbb{R}^{m \times n}$ that has rank no greater than $k$, 
$B \in \mathbb{R}^{m \times p}$,
sampling probabilities $\{p_i\}_{i=1}^m$, and 
$r \leq m$.
}

\AlgResult{
$\tilde{X}_{opt} \in \mathbb{R}^{n \times p}$, 
$\tilde{\cal Z} \in \mathbb{R}$.
}

\begin{itemize}
\item
(Implicitly) construct a sampling matrix $S$ and a diagonal rescaling 
matrix $D$ with the \textsc{Exactly($c$)} algorithm or with the 
\textsc{Expected($c$)} algorithm\;
\item
Construct the matrix $DS^TA$ consisting of a small number of rescaled rows 
of $A$\;
\item
Construct the matrix $DS^TB$ consisting of a small number of rescaled rows 
of $B$\;
\item
$\tilde{X}_{opt} = \left(DS^TA\right)^+ DS^TB$\;
\item
$\tilde{\cal Z} = \min_{X \in \mathbb{R}^{n \times p}} \FNorm{DS^TB - DS^TA\tilde{X}_{opt}}$\;
\end{itemize}

\end{framed}

\caption{A Monte-Carlo algorithm for approximating $\ell_2$ regression.}

\label{alg:sample_l2regression}

\end{algorithm}


Theorem~\ref{thm:ls_bound} is our main quality-of-approximation result for 
Algorithm~\ref{alg:sample_l2regression}.
Its proof may be found in Section~\ref{sxn:proof}.
Recall that for our generalized $\ell_2$ regression problem, the matrix 
$A$ has rank no greater than $k$.

\begin{theorem}
\label{thm:ls_bound}
Suppose $A \in \mathbb{R}^{m \times n}$ has rank no greater than $k$, 
$B \in \mathbb{R}^{m \times p}$,
$\epsilon \in (0,1]$, and let
$
{\cal Z} 
   = \min_{X \in \mathbb{R}^{n \times p}} \FNorm{B - AX} 
   = \FNorm{B-AX_{opt}}     
$,
where $X_{opt}=A^+B=A_k^+B$.
Run Algorithm~\ref{alg:sample_l2regression} with any sampling probabilities
of the form
\begin{equation}
\label{eqn:ass1}
p_i \geq \beta
         \frac{             \VTTNormS{\left(U_{A,k}\right)_{(i)}}}
              {\sum_{j=1}^n \VTTNormS{\left(U_{A,k}\right)_{(j)}}}     
       = \frac{\beta}{k} \VTTNormS{\left(U_{A,k}\right)_{(i)}}          ,
   \hspace{0.25in}
   \mbox{   }
   \forall i \in [n]   ,
\end{equation}
for some $\beta \in (0,1]$, and assume that the output of the algorithm is a 
number $\tilde{\cal Z}$ and an $n \times p$ matrix $\tilde{X}_{opt}$.
If exactly $ r = 3200k^2/\beta\epsilon^2$ rows are chosen with the 
\textsc{Exactly($c$)} algorithm, then with probability at least $0.7$:
\begin{eqnarray}
\label{eqn:result2}
\FNorm{B - A\tilde{X}_{opt}}
   &\leq& \left( 1+\epsilon \right){\cal Z},           \\
\label{eqn:result3}
\FNorm{X_{opt} - \tilde{X}_{opt}}
   &\leq& \frac{\epsilon}{\sigma_{\min}(A_k)} {\cal Z}     .
\end{eqnarray}
If, in addition, we assume that 
$\FNorm{U_{A,k}U_{A,k}^TB} \geq \gamma \FNorm{B} $,
for some fixed $\gamma \in (0,1]$, then with probability at least $0.7$:
\begin{equation}
\label{eqn:result4}
\FNorm{X_{opt} - \tilde{X}_{opt}}
   \leq \epsilon \left( \kappa(A_k)\sqrt{\gamma^{-2}-1} \right) \FNorm{X_{opt}}     .
\end{equation}
Similarly, under the same assumptions, if $r=O(k \log k/\beta\epsilon^2)$ rows 
are chosen in expectation with the \textsc{Expected($c$)} algorithm, then with 
probability at least $0.7$,  
(\ref{eqn:result2}), (\ref{eqn:result3}), and (\ref{eqn:result4}) hold.
\end{theorem}

Equation (\ref{eqn:result2}) states that if the matrix of 
minimum-length vectors achieving the minimum in the sampled problem is 
substituted back into the residual norm for the original problem then a good approximation to 
the original $\ell_2$ regression problem is obtained. 
Equation (\ref{eqn:result3}) provides a bound for 
$\FNorm{X_{opt} - \tilde{X}_{opt}}$ in terms of $\sigma_{\min}(A_k)$ and 
${\cal Z}$. 
If most of the ``weight'' of $B$ lies in the complement of the column space 
of $A=A_k$ then this will provide a very poor approximation in terms of 
$\FNorm{X_{opt}}$. 
However, if we also assume that a constant fraction of the ``weight'' of $B$ 
lies in the subspace spanned by the columns of $A$, then we obtain the 
relative-error approximation of Equation~(\ref{eqn:result4}). 
Thus, Theorem~\ref{thm:ls_bound} returns a good bound for 
$\FNorm{X_{opt} - \tilde{X}_{opt}}$ if $A_k$ is well-conditioned and if $B$ 
lies ``reasonably well'' in the column space of $A$.
Note that if the matrix of target vectors $B$ lies completely within the 
column space of $A$, then $\mathcal{Z}=0$ and $\gamma=1$.
In this case, Theorem~\ref{thm:ls_bound} shows that 
Algorithm~\ref{alg:sample_l2regression} returns $\tilde{\mathcal{Z}}$ and 
$\tilde{x}_{opt}$ that are exact solutions of the original $\ell_2$ regression 
problem, independent of $\kappa(A_k)$.
Finally, note that in our analysis of CX and CUR matrix decompositions
we only use the result (\ref{eqn:result2}) from Theorem~\ref{thm:ls_bound},
but (\ref{eqn:result3}) and (\ref{eqn:result4}) are included for completeness.

\subsection{Discussion of the Method of ``Subspace Sampling''}
\label{sxn:main_l2_alg:discussion}

An important aspect of Algorithm~\ref{alg:sample_l2regression} is the 
nonuniform sampling probabilities (\ref{eqn:ass1}) used by the 
\textsc{Exactly($c$)} algorithm and the \textsc{Expected($c$)} algorithm in 
the construction of the induced subproblem.
We call sampling probabilities satisfying condition (\ref{eqn:ass1}) 
\emph{``subspace sampling''} probabilities.
Condition (\ref{eqn:ass1}) states that the sampling probabilities should be 
close to, or rather not much less than, the lengths, i.e., the Euclidean 
norms, of the rows of the left singular vectors of the matrix $A=A_k$.
(Recall that in this section $A$ is an $m \times n$ matrix with rank no more 
than $k$, and thus $U_{A,k}$ is an $m \times k$ matrix.
Thus, the Euclidean norm of every \emph{column} of $U_{A,k}$ equals $1$, but 
the Euclidean norm of every \emph{row} of $U_{A,k}$ is in general not equal 
and is only bounded above by $1$.)
Sampling probabilities of the form (\ref{eqn:ass1})
should be contrasted with sampling probabilities that depend on the 
Euclidean norms of the columns or rows of $A$ and that have received much 
attention recently~\cite{FKV98,FKV04,dkm_matrix1,dkm_matrix2,dkm_matrix3,dm_tensorSVD_JRNL}.
Since $A=U_A\Sigma_AV_A^T$, sampling probabilities with this latter form 
depend in a complicated manner on a mixture of subspace information (as found 
in $U_A$ and $V_A$) and 
``size-of-$A$'' information (as found in $\Sigma_A$).
This convolution of information may account for their ability to capture 
coarse statistics such as approximating matrix multiplication or computing 
low-rank matrix approximations to additive error, but it also accounts for 
their difficulty in dealing with problems such as $\ell_2$ regression or 
computing low-rank matrix approximations to relative error.

Since the solution of the $\ell_2$ regression problem involves the computation 
of a pseudoinverse, the problem is not well-conditioned with respect to a 
perturbation (such as that introduced by sampling) that entails a change in 
dimensionality, even if (actually, especially if) that change in dimensionality 
corresponds to a small singular value.
Since sampling probabilities satisfying (\ref{eqn:ass1}) allow us to 
disentangle subspace information and ``size-of-$A$'' information, we will see 
that they will allow us to capture (with high probability) the \emph{entire} 
subspace of interest by sampling.
More precisely, as we will see in Lemma~\ref{lemma:matmult1}, by using sampling 
probabilities that satisfy condition (\ref{eqn:ass1}) and by choosing $r$
appropriately, it will follow that
$$
\mbox{rank}(DS^TU_{A,k}) = \mbox{rank}(U_{A,k}) = k   .
$$
Thus, the lengths of the Euclidean norms of the rows of $U_{A,k}$ may be 
interpreted as capturing a notion of information dispersal by the matrix $A$ 
since they indicate to which part of the $m$-dimensional 
vector space the singular value information of $A$ is being dispersed.
In this case, condition (\ref{eqn:ass1}) ensures that the sampling 
probabilities provide a bias toward the part of the high-dimensional 
constraint space to which $A$ disperses its singular value information.
Then, having constructed the sample, we will go to the low-dimensional, i.e., 
the $r$-dimensional rather than then $m$-dimensional space, and approximate 
the $\ell_2$ regression problem by doing computations that involve 
``size-of-$A$'' information on the random sample. 

This method of ``subspace sampling'' was first used in a preliminary version
of the $\ell_2$ regression results of this section \cite{DMM06}.
Note that an immediate generalization of the results of \cite{DMM06} to the 
generalized $\ell_2$ regression problem considered in this section would 
involve sampling probabilities of the form
\begin{equation}
\label{eqn:probs_leastsquares}
p_i =
   \frac{ (1/3)       \VTTNormS{\left(U_{A,k}\right)_{(i)}}}
        {\sum_{j=1}^n \VTTNormS{\left(U_{A,k}\right)_{(j)}}}          
   +
   \frac{ (1/3)       \VTTNorm{\left(U_{A,k}\right)_{(i)}}{\left(U_{A,k}^{\perp}{U_{A,k}^{\perp}}^TB\right)_i}}
        {\sum_{j=1}^n \VTTNorm{\left(U_{A,k}\right)_{(j)}}{\left(U_{A,k}^{\perp}{U_{A,k}^{\perp}}^TB\right)_j}}          
   +
   \frac{ (1/3)       \left(U_{A,k}^{\perp}{U_{A,k}^{\perp}}^TB\right)_i^2}
        {\sum_{j=1}^n \left(U_{A,k}^{\perp}{U_{A,k}^{\perp}}^TB\right)_j^2}   ,
\end{equation}
rather than of the form (\ref{eqn:ass1}).
Since the second and third terms in (\ref{eqn:probs_leastsquares}) provide a 
bias toward the part of the complement of the column space of $A=A_k$ where 
$B$ has significant weight, we directly obtain variance reduction.
Thus, by using probabilities of the form (\ref{eqn:probs_leastsquares}) we 
can sample $O(k^2 \log(1/\delta)/\epsilon^2)$ columns and directly obtain the 
claims of Theorem~\ref{thm:ls_bound} with probability at least $1-\delta$.
Although sampling probabilities of the form (\ref{eqn:ass1}) are substantially 
simpler, we obtain variance control indirectly.
We first establish that each of the claims of Theorem~\ref{thm:ls_bound} holds 
with constant probability, and we then can show that each of the claims holds 
with probability at least $1-\delta$ by running $O(\log(1/\delta))$ trials and 
using standard boosting procedures.

\subsection{Proof of Theorem \ref{thm:ls_bound}}
\label{sxn:proof}

In this section we provide a proof of Theorem~\ref{thm:ls_bound}.
We will first prove (\ref{eqn:result2}), (\ref{eqn:result3}), and
(\ref{eqn:result4}) under the assumption that the rows of $A$ and $B$ are 
sampled with the \textsc{Exactly($c$)} algorithm.
Then, in Section~\ref{sxn:proof_modifications}, we will outline modifications 
to the proof if the rows of $A$ and $B$ are sampled with the 
\textsc{Expected($c$)} algorithm.
For simplicity of notation in this section, we will let $\mathcal{S}=DS^T$ 
denote the $r \times m$ rescaled row-sampling matrix.
Let the rank of the $m \times n$ matrix $A$ be $\rho \leq k$, and let its SVD be
$$
A = U_A \Sigma_A V_A^T      ,
$$ 
where $U_A \in \mathbb{R}^{n \times \rho}$, 
$\Sigma_A \in \mathbb{R}^{\rho \times \rho}$, and
$V_A \in \mathbb{R}^{d \times \rho}$. 
In addition, let the rank of the $r \times \rho$ matrix 
$\mathcal{S}U_A = DS^TU_A$ be $\tilde{\rho}$, and let its SVD be
$$
\mathcal{S}U_A 
   = U_{\mathcal{S}U_A} \Sigma_{\mathcal{S}U_A} V_{\mathcal{S}U_A}^T     ,
$$ 
where $U_{\mathcal{S}U_A} \in \mathbb{R}^{r \times \tilde{\rho}}$, 
$\Sigma_{\mathcal{S}U_A} \in \mathbb{R}^{\tilde{\rho} \times \tilde{\rho}}$, 
and $V_{\mathcal{S}U_A} \in \mathbb{R}^{\rho \times \tilde{\rho}}$.
Recall that $\tilde{\rho} \le \rho \le k \le r$.

In order to illustrate the essential difficulty in constructing a sampling 
algorithm to approximate the solution of the generalized $\ell_2$ regression 
problem, consider inserting $\tilde{X}_{opt} = (\mathcal{S}A_k)^+\mathcal{S}B$ 
into $B - A_kX$: 
\begin{eqnarray*}
B - A_k\tilde{X}_{opt} 
   &=& B - A_k \left(\mathcal{S}A_k\right)^+\mathcal{S}B          \\
   &=& B - U_{A,k}\Sigma_{A,k}V_{A,k}^T \left(\mathcal{S}U_{A,k}\Sigma_{A,k}V_{A,k}^T\right)^+\mathcal{S}B     \\
   &=& B - U_{A,k}\Sigma_{A,k} \left(\mathcal{S}U_{A,k}\Sigma_{A,k}\right)^+\mathcal{S}B     \\
   &=& B - U_{A,k}\Sigma_{A,k} \left( U_{\mathcal{S}U_{A,k}} \Sigma_{\mathcal{S}U_{A,k}} V_{\mathcal{S}U_{A,k}}^T \Sigma_{A,k}\right)^+\mathcal{S}B     \\
   &=& B - U_{A,k}\Sigma_{A,k} \left( \Sigma_{\mathcal{S}U_{A,k}} V_{\mathcal{S}U_{A,k}}^T \Sigma_{A,k}\right)^+ U_{\mathcal{S}U_{A,k}}^T \mathcal{S}B     .
\end{eqnarray*}
To proceed further, we must deal with the pseudoinverse, which is not 
well-behaved with respect to perturbations that involve a change in 
dimensionality.
To deal with this, we will focus on probabilities that depend on the subspace 
that we are downsampling, i.e., that depend on $U_{A,k}$, in order to guarantee 
that we capture the full subspace of interest.

\subsubsection{Several lemmas of general interest}
\label{sxn:proof:lemmas}

In this subsection, we will present three lemmas of general interest.
Then, in the next subsections, we will use these lemmas to prove each of the 
claims of Theorem~\ref{thm:ls_bound}.

Since the $m \times k$ matrix $U_{A,k}$ is a matrix with orthogonal columns,
several properties hold for it.
For example, $\mbox{rank}(U_{A,k}) = k$, $U^+=U^T$, and 
$A_k^+ = V_{A,k}\Sigma_{A,k}^{-1}U_{A,k}^T$.
Although the $r \times k$ matrix $\mathcal{S}U_{A,k}$ is does not have 
orthogonal columns, the following lemma characterizes the manner in which 
each of these three properties holds, either exactly or approximately.
For the first lemma, $r$ depends quadratically on $k$.

\begin{lemma} 
\label{lemma:matmult1}
Let $\epsilon \in (0,1]$, and define
$\Omega=\left(\mathcal{S}U_{A,k}\right)^+ - \left(\mathcal{S}U_{A,k}\right)^T$.
If the sampling probabilities satisfy equation (\ref{eqn:ass1}) and if 
$r \geq 400 k^2/\beta\epsilon^2$, then with probability at least $0.9$:
\begin{eqnarray}
& & \tilde{\rho} 
    =    \rho \mbox{, i.e., }
         \mbox{rank}(\mathcal{S}U_{A,k}) = \mbox{rank}(U_{A,k}) = \mbox{rank}(A_k)   \\
& &\TNorm{\Omega} 
   = \TNorm{\Sigma_{\mathcal{S}U_{A,k}}^{-1} - \Sigma_{\mathcal{S}U_{A,k}}}   \\
\label{eqn3:lemma:matmult1}
& & \left(\mathcal{S}A_k\right)^+ 
    =    V_{A,k}\Sigma_{A,k}^{-1}\left(\mathcal{S}U_{A,k}\right)^+            \\
& & \TNorm{\Sigma_{\mathcal{S}U_{A,k}} - \Sigma_{\mathcal{S}U_{A,k}}^{-1}} 
    \leq \epsilon/\sqrt{2}                                                  
\end{eqnarray}
\end{lemma}
\begin{Proof}
To prove the first claim, note that for all $i \in [\rho]$
\begin{eqnarray}
\abs{1 - \sigma_i^2\left(\mathcal{S}U_{A,k}\right)} 
\nonumber
   &=&    \abs{\sigma_i\left(U_{A,k}^TU_{A,k}\right) 
            - \sigma_i\left(U_{A,k}^T\mathcal{S}^T\mathcal{S}U_{A,k}\right)}  \\
\label{eq18}
   &\leq& \TNorm{U_{A,k}^TU_{A,k} - U_{A,k}^T\mathcal{S}^T\mathcal{S}U_{A,k}} \\
\label{eq19} 
   &\leq& \FNorm{U_{A,k}^TU_{A,k} - U_{A,k}^T\mathcal{S}^T\mathcal{S}U_{A,k}}  .
\end{eqnarray}
Note that (\ref{eq18}) follows from Corollary 8.1.6 of \cite{GVL96}, and
(\ref{eq19}) follows since $\TNorm{\cdot}\le\FNorm{\cdot}$.
To bound the error of approximating $U_{A,k}^T U_{A,k}$ by 
$U_{A,k}^T\mathcal{S}^T\mathcal{S}U_{A,k}$ we apply 
Theorem~\ref{thm:matmult-main-exact} of Appendix~\ref{sxn:matrix_multiply}.
Since the sampling probabilities $p_i$ satisfy equation (\ref{eqn:ass1}), it 
follows from Theorem~\ref{thm:matmult-main-exact} and by applying Markov's 
inequality that with probability at least $0.9$:
\begin{eqnarray}
\FNorm{U_{A,k}^TU_{A,k} - U_{A,k}^T\mathcal{S}^T\mathcal{S}U_{A,k}}           
\nonumber
   &\leq& 10 \; \Expect{\FNorm{U_{A,k}^TU_{A,k} - U_{A,k}^T\mathcal{S}^T\mathcal{S}U_{A,k}}}    \\
\label{eqn:tt1} 
   &\leq& \frac{10}{\sqrt{\beta r}}\FNormS{U_{A,k}}    ,
\end{eqnarray}
where $\Expect{\cdot}$ denotes the expectation operator.
By combining (\ref{eq19}) and (\ref{eqn:tt1}), 
recalling that $\FNormS{U_{A,k}} = \rho \leq k$,
and using the assumed choice of $r$, it follows that
$$
\abs{1 - \sigma_i^2\left(\mathcal{S}U_{A,k}\right)} 
   \leq \epsilon/2 
   \leq 1/2 
$$
since $\epsilon \le 1$.
This implies that all singular values of $\mathcal{S}U_{A,k}$ are strictly 
positive, and thus that
$\mbox{rank}(\mathcal{S}U_{A,k}) = \mbox{rank}(U_{A,k}) = \mbox{rank}(A_k)$, 
which establishes the first claim.

To prove the second claim, we use the SVD of $\mathcal{S}U_{A,k}$ and note that
\begin{eqnarray*}
\TNorm{\Omega} 
  &=& \TNorm{\left(\mathcal{S}U_{A,k}\right)^+ - \left(\mathcal{S}U_{A,k}\right)^T}  \\
  &=& \TNorm{\left(U_{\mathcal{S}U_{A,k}}\Sigma_{\mathcal{S}U_{A,k}}V_{\mathcal{S}U_{A,k}}^T\right)^+ - \left(U_{\mathcal{S}U_{A,k}}\Sigma_{\mathcal{S}U_{A,k}}V_{\mathcal{S}U_{A,k}}^T\right)^T}  \\
  &=& \TNorm{V_{\mathcal{S}U_{A,k}}\left(\Sigma_{\mathcal{S}U_{A,k}}^{-1} - \Sigma_{\mathcal{S}U_{A,k}}\right)U_{\mathcal{S}U_{A,k}}^T}   .
\end{eqnarray*}
The claim follows since $V_{\mathcal{S}U_{A,k}}$ and $U_{\mathcal{S}U_{A,k}}$ 
are matrices with orthonormal columns.

To prove the third claim, note that
\begin{eqnarray}
\nonumber
\left(\mathcal{S}A_k\right)^+ 
   &=& \left(\mathcal{S}U_{A,k} \Sigma_{A,k} V_{A,k}^T\right)^+          \\
\nonumber
   &=& \left(U_{\mathcal{S}U_{A,k}} \Sigma_{\mathcal{S}U_{A,k}}V_{\mathcal{S}U_{A,k}}^T\Sigma_{A,k}V_{A,k}^T\right)^+     \\
\label{eq20} 
   &=& V_{A,k}\left(\Sigma_{\mathcal{S}U_{A,k}} V_{\mathcal{S}U_{A,k}}^T\Sigma_{A,k}\right)^+ U_{\mathcal{S}U_{A,k}}^T     .
\end{eqnarray}
To remove the pseudoinverse in the above derivations, notice that since 
$\rho = \tilde{\rho}$ with probability at least $0.9$, all three matrices 
$\Sigma_{\mathcal{S}U_{A,k}}$, $V_{\mathcal{S}U_{A,k}}$, and $\Sigma_{A,k}$ 
are full rank square $\rho \times \rho$ matrices, and thus are invertible. 
In this case,
\begin{eqnarray}
\nonumber
\left(\Sigma_{\mathcal{S}U_{A,k}} V_{\mathcal{S}U_{A,k}}^T\Sigma_{A,k}\right)^+ 
   &=& \left(\Sigma_{\mathcal{S}U_{A,k}} V_{\mathcal{S}U_{A,k}}^T\Sigma_{A,k}\right)^{-1} \\
\label{eq201} 
   &=& \Sigma_{A,k}^{-1} V_{\mathcal{S}U_{A,k}} \Sigma_{\mathcal{S}U_{A,k}}^{-1}           .
\end{eqnarray}
By combining (\ref{eq20}) and (\ref{eq201}) we have that
\begin{eqnarray*}
\left(\mathcal{S}A_k\right)^+ 
   &=&  V_{A,k} \Sigma_{A,k}^{-1} V_{\mathcal{S}U_{A,k}} \Sigma_{\mathcal{S}U_{A,k}}^{-1} U_{\mathcal{S}U_{A,k}}^T      \\
   &=& V_{A,k} \Sigma_{A,k}^{-1} \left(\mathcal{S}U_{A,k}\right)^+     ,
\end{eqnarray*}
which establishes the third claim.%
\footnote{One might be tempted to suggest that the proof of this third claim 
should be ``simplified'' by appealing to the result that the generalized 
inverse of the product of two matrices equals the product of the generalized 
inverse of those matrices.  This result is, of course, false---see, e.g., 
Section 3.1.1 of \cite{Stewart90}--and so we need a more refined analysis 
such as the one presented here.}

Finally, to prove the fourth claim, recall that under the assumptions of the 
lemma $\rho = \tilde{\rho}$ with probability at least $0.9$, and thus 
$\sigma_i\left(\mathcal{S}U_{A,k}\right) > 0$ for all $i \in [\rho]$.
Thus,
\begin{eqnarray}
\TNorm{\Sigma_{\mathcal{S}U_{A,k}}^{-1} - \Sigma_{\mathcal{S}U_{A,k}}} 
\nonumber
   &=&    \max_{i,j \in [\rho]}\abs{\sigma_i\left(\mathcal{S}U_{A,k}\right) - \frac{1}{\sigma_j\left(\mathcal{S}U_{A,k}\right)}}          \\
\nonumber
   &=&    \max_{i,j \in [\rho]}\frac{\abs{\sigma_i\left(\mathcal{S}U_{A,k}\right)\sigma_j\left(\mathcal{S}U_{A,k}\right) - 1}}{\abs{\sigma_j\left(\mathcal{S}U_{A,k}\right)}}\\
\label{eqn:lem1pfeqn1}
   &\leq& \max_{j \in [\rho]}\frac{\abs{\sigma_j^2\left(\mathcal{S}U_{A,k}\right) - 1}}{\abs{\sigma_j\left(\mathcal{S}U_{A,k}\right)}}            .
\end{eqnarray}
Using that fact that, by (\ref{eq18}), for all $i \in [\rho]$, 
\begin{equation*}
\abs{1 - \sigma_i^2\left(\mathcal{S}U_{A,k}\right)} 
   \leq \TNorm{U_{A,k}^TU_{A,k} - U_{A,k}^T\mathcal{S}^T\mathcal{S}U_{A,k}}   , 
\end{equation*}
it follows that for all $i \in [\rho]$
\begin{equation*}
\frac{1}{\sigma_i\left(\mathcal{S}U_{A,k}\right)} 
   \leq \frac{1}{\sqrt{1-\TNorm{U_{A,k}^TU_{A,k} - U_{A,k}^T\mathcal{S}^T\mathcal{S}U_{A,k}}}} .
\end{equation*}
When these are combined with (\ref{eqn:lem1pfeqn1}) it follows that
\begin{equation*}
\TNorm{\Sigma_{\mathcal{S}U_{A,k}} - \Sigma_{\mathcal{S}U_{A,k}}^{-1}} 
   \leq \frac{\TNorm{U_{A,k}^TU_{A,k} - U_{A,k}^T\mathcal{S}^T\mathcal{S}U_{A,k}}}
             {\sqrt{1-\TNorm{U_{A,k}^TU_{A,k} - U_{A,k}^T\mathcal{S}^T\mathcal{S}U_{A,k}}}}    .
\end{equation*}
Combining this with the Frobenius norm bound of (\ref{eqn:tt1}), and noticing 
that our choice for $r$ guarantees that 
$1 - \TNorm{U_{A,k}^TU_{A,k} - U_{A,k}^T\mathcal{S}^T\mathcal{S}U_{A,k}} \geq 1/2$, concludes 
the proof of the fourth claim.

This concludes the proof of the lemma.
\end{Proof}

The next lemma provides an approximate matrix multiplication bound that is 
useful in the proof of Theorem~\ref{thm:ls_bound}.
For this lemma, $r$ depends linearly on $k$.

\begin{lemma} 
\label{lemma:matmult2}
Let $\epsilon \in (0,1]$. 
If the sampling probabilities satisfy equation (\ref{eqn:ass1}) and if 
$r \geq 400k/\beta\epsilon^2$, then with probability at least $0.9$:
$$
\FNorm{U_{A,k}^T\mathcal{S}^T\mathcal{S}U_{A,k}^{\perp}{U_{A,k}^{\perp}}^T B}  
   \leq \frac{\epsilon}{2} \FNorm{U_{A,k}^{\perp} {U_{A,k}^{\perp}}^T B}     .
$$
\end{lemma}
\begin{Proof}
First, note that since $U_{A,k}$ is an orthogonal matrix and since 
$ U_{A,k}^TU_{A,k}^{\perp} = 0 $ , we have that
\begin{eqnarray}
\FNorm{U_{A,k}^T\mathcal{S}^T \mathcal{S} U_{A,k}^{\perp} {U_{A,k}^{\perp}}^T B}  
\nonumber
   &=& \FNorm{U_{A,k}U_{A,k}^T\mathcal{S}^T \mathcal{S}U_{A,k}^{\perp}{U_{A,k}^{\perp}}^T B}  \\
\label{eqn:lem_mm3_eq1}
   &=& \FNorm{
        U_{A,k} U_{A,k}^T U_{A,k}^{\perp} {U_{A,k}^{\perp}}^T B
        - U_{A,k} U_{A,k}^T \mathcal{S}^T \mathcal{S} U_{A,k}^{\perp} {U_{A,k}^{\perp}}^T B}     .
\end{eqnarray}
Since
$\VTTNorm{\left(U_{A,k}U_{A,k}^T\right)_{(i)}} = \VTTNorm{\left(U_{A,k}^T\right)_{(i)}}$, 
the sampling probabilities (\ref{eqn:ass1}) satisfy 
(\ref{eqn:nonoptimal_probs}), where (\ref{eqn:nonoptimal_probs}) will appear 
in Appendix~\ref{sxn:matrix_multiply:alg},
and thus are appropriate for bounding the right 
hand side of (\ref{eqn:lem_mm3_eq1}).
Thus, it follows from 
Markov's inequality and Theorem~\ref{thm:matmult-main-exact} 
that with probability at least $0.9$:
\begin{eqnarray*}
\FNorm{U_{A,k}^T\mathcal{S}^T \mathcal{S}U_{A,k}^{\perp}{U_{A,k}^{\perp}}^T B}  
   &\leq& 10 \; \Expect{ \FNorm{U_{A,k}^T\mathcal{S}^T \mathcal{S}U_{A,k}^{\perp}{U_{A,k}^{\perp}}^T B} } \\
   &\leq& \frac{10}{\sqrt{\beta r}} 
          \FNorm{U_{A,k}U_{A,k}^T} 
          \FNorm{U_{A,k}^{\perp} {U_{A,k}^{\perp}}^T B}      .
\end{eqnarray*}
The lemma follows by the choice of $r$ and since 
$\FNorm{U_{A,k}U_{A,k}^T} = \sqrt{\rho} \leq \sqrt{k}$.
\end{Proof}

The final lemma of this subsection relates the norm of the $m \times p$ matrix
$U_{A,k}^{\perp}{U_{A,k}^{\perp}}^TB$ to the norm of the $r \times p$ matrix
$\mathcal{S}U_{A,k}^{\perp}{U_{A,k}^{\perp}}^TB$, i.e., the row sampled and 
rescaled version of the original $m \times p$ matrix.
For this lemma, $r$ is independent of $k$.

\begin{lemma} 
\label{lemma:matmult3}
With probability at least $0.9$:
$$
\FNorm{\mathcal{S}U_{A,k}^{\perp}{U_{A,k}^{\perp}}^TB}   
   \leq 10 \FNorm{U_{A,k}^{\perp}{U_{A,k}^{\perp}}^TB}     .
$$
\end{lemma}
\begin{Proof}
Let $Q = U_{A,k}^{\perp}{U_{A,k}^{\perp}}^TB$, and let $j_1, j_2,\ldots,j_r$ 
be the $r$ rows of $Q$ that were included in $\mathcal{S}Q = D S^T Q$. 
Clearly,
\begin{eqnarray}
\Expect{\FNormS{DS^TQ}} 
   = \Expect{\sum_{t=1}^r \VTTNormS{Q_{(j_t)}}}
   = \sum_{t=1}^r \Expect{\VTTNormS{Q_{(j_t)}}} 
   = \sum_{t=1}^r \sum_{j=1}^n p_j \frac{\VTTNormS{Q_{(j)}}}{rp_j} 
   = \FNormS{Q}    ,
\end{eqnarray}
where the penultimate equality follows by evaluating the expectation.
The lemma follows by applying Markov's inequality and taking the square root 
of both sides of the resulting inequality.
\end{Proof}

\subsubsection{Proof of Equation (\ref{eqn:result2})}
\label{sxn:result2}

In this subsection, we will bound $B-A_k\tilde{X}_{opt}$, thus proving 
(\ref{eqn:result2}).
For the moment, let us assume that $r = 400 k^2/ \beta \epsilon^2$, in which 
case the assumption on $r$ is satisfied for each of
Lemma~\ref{lemma:matmult1}, Lemma~\ref{lemma:matmult2}, and 
Lemma~\ref{lemma:matmult3}.
Thus, the claims of all three lemmas hold simultaneously with probability at 
least $1-3(0.1) \geq 0.7$, and so let us condition on this event.

First, we have that
\begin{eqnarray}
B - A_k\tilde{X}_{opt} 
\nonumber
   &=& B - A_k \left(\mathcal{S}A_k\right)^+\mathcal{S}B                     \\
\label{eqn:pfBeq1}
   &=& B - U_{A,k} \left(\mathcal{S}U_{A,k}\right)^+\mathcal{S}B             \\
\label{eqn:pfBeq2}
   &=& B - U_{A,k} 
       \left(\mathcal{S}U_{A,k}\right)^+ \mathcal{S} U_{A,k}
       U_{A,k}^T B     
    - U_{A,k} \left(\mathcal{S}U_{A,k}\right)^+ \mathcal{S} U_{A,k}^{\perp} {U_{A,k}^{\perp}}^T B     \\
\label{eqn:pfBeq3}
   &=& U_{A,k}^{\perp} {U_{A,k}^{\perp}}^T B - U_{A,k} \left(\mathcal{S}U_{A,k}\right)^+ \mathcal{S} U_{A,k}^{\perp} {U_{A,k}^{\perp}}^T B     .
\end{eqnarray}
(\ref{eqn:pfBeq1}) follows from (\ref{eqn3:lemma:matmult1}) of Lemma~\ref{lemma:matmult1},
(\ref{eqn:pfBeq2}) follows by inserting 
$ U_{A,k}U_{A,k}^T + U_{A,k}^{\perp}{U_{A,k}^{\perp}}^T = I_n$, and
(\ref{eqn:pfBeq3}) follows since 
$\left(\mathcal{S}U_{A,k}\right)^+ \mathcal{S} U_{A,k} = I_{\rho}$ by
Lemma~\ref{lemma:matmult1}. 
We emphasize that
$
\left(\mathcal{S}U_{A,k}\right)^+ \mathcal{S} U_{A,k} 
   = V_{\mathcal{S}U_{A,k}} V_{\mathcal{S}U_{A,k}}^T = I_{\rho}
$
does not hold for general sampling methods, but it does hold in this case since 
$\tilde{\rho}=\rho$, which follows from Lemma~\ref{lemma:matmult1}.

By taking the Frobenius norm of both sides of (\ref{eqn:pfBeq3}), by using the 
triangle inequality, and recalling that 
$\Omega= \left(\mathcal{S}U_{A,k}\right)^+ - \left(\mathcal{S}U_{A,k}\right)^T$, we 
have that
\begin{eqnarray}
\FNorm{B - A_k\tilde{X}_{opt}} 
\nonumber
   &\leq& 
          \FNorm{U_{A,k}^{\perp} {U_{A,k}^{\perp}}^T B} 
        + \FNorm{U_{A,k} \left(\mathcal{S}U_{A,k}\right)^T \mathcal{S} U_{A,k}^{\perp} {U_{A,k}^{\perp}}^T B}     
        + \FNorm{U_{A,k} \Omega \mathcal{S} U_{A,k}^{\perp} {U_{A,k}^{\perp}}^T B}      \\
\label{eqn:pfBeq4}
   &\leq& 
          \FNorm{U_{A,k}^{\perp} {U_{A,k}^{\perp}}^T B} 
        + \FNorm{U_{A,k}^T\mathcal{S}^T \mathcal{S}U_{A,k}^{\perp}{U_{A,k}^{\perp}}^T B}
        + \TNorm{\Omega}\FNorm{\mathcal{S} U_{A,k}^{\perp} {U_{A,k}^{\perp}}^T B}    ,
\end{eqnarray}
where (\ref{eqn:pfBeq4}) follows by submultiplicativity and since $U_{A,k}$ 
has orthogonal columns.
By combining (\ref{eqn:pfBeq4}) with the bounds provided by 
Lemma~\ref{lemma:matmult1} through Lemma~\ref{lemma:matmult3}, it follows that
\begin{eqnarray*}
\FNorm{B - A_k\tilde{X}_{opt}}
   &\leq& (1+\epsilon/2+10\epsilon/\sqrt{2})\mathcal{Z}  \\
   &\leq& (1+ 8 \epsilon)\mathcal{Z}                                        .
\end{eqnarray*}
Equation (\ref{eqn:result2}) follows by setting $\epsilon' = \epsilon/8$ and 
using the value of $r$ assumed by the theorem.

\subsubsection{Proof of Equation (\ref{eqn:result3})}

In this subsection, we will provide a bound for 
$\FNorm{\tilde{X}_{opt} - X_{opt}}$ in terms of $\mathcal{Z}$, thus proving 
(\ref{eqn:result3}).
For the moment, let us assume that $r = 400 k^2/\beta \epsilon^2$, in which 
case the assumption on $r$ is satisfied for each of
Lemma~\ref{lemma:matmult1}, Lemma~\ref{lemma:matmult2}, and 
Lemma~\ref{lemma:matmult3}.
Thus, the claims of all three lemmas hold simultaneously with probability at 
least $1-3(0.1) \geq 0.7$, and so let us condition on this event.

Since $U_{A,k}U_{A,k}^T+U_{A,k}^{\perp}{U_{A,k}^{\perp}}^T = I_n$ and
$\left(\mathcal{S}U_{A,k}\right)^+\mathcal{S}U_{A,k} = I_{\rho}$, we have that
\begin{eqnarray*}
X_{opt}-\tilde{X}_{opt}
 &=&    A_k^+B - \left(\mathcal{S}A_k\right)^+\mathcal{S}B                    \\
 &=&    V_{A,k} \Sigma_{A,k}^{-1} U_{A,k}^T B  
      - V_{A,k}\Sigma_{A,k}^{-1}\left(\mathcal{S}U_{A,k}\right)^+\mathcal{S}B         \\
 &=&    V_{A,k} \Sigma_{A,k}^{-1} U_{A,k}^T B 
      - V_{A,k}\Sigma_{A,k}^{-1}\left(\mathcal{S}U_{A,k}\right)^+\mathcal{S}U_{A,k}U_{A,k}^TB  
      - V_{A,k}\Sigma_{A,k}^{-1}\left(\mathcal{S}U_{A,k}\right)^+\mathcal{S}U_{A,k}^{\perp}{U_{A,k}^{\perp}}^TB  \\
 &=&  - V_{A,k}\Sigma_{A,k}^{-1}\left(\mathcal{S}U_{A,k}\right)^+\mathcal{S}U_{A,k}^{\perp}{U_{A,k}^{\perp}}^TB   .
\end{eqnarray*}
Thus, it follows that
\begin{eqnarray}
\FNorm{X_{opt}-\tilde{X}_{opt}}
\nonumber
   &=&     \FNorm{ V_{A,k}\Sigma_{A,k}^{-1}
                     \left(\mathcal{S}U_{A,k}\right)^+
                     \mathcal{S}U_{A,k}^{\perp}{U_{A,k}^{\perp}}^TB }                \\
\nonumber
   &=&     \FNorm{ \Sigma_{A,k}^{-1}
                     \left(\left(\mathcal{S}U_{A,k}\right)^T+\Omega\right)
                     \mathcal{S}U_{A,k}^{\perp}{U_{A,k}^{\perp}}^TB }                \\
\nonumber
   &\leq&  \frac{1}{\sigma_{min}(A_k)}
           \FNorm{ \left(\mathcal{S}U_{A,k}\right)^T \mathcal{S}U_{A,k}^{\perp}{U_{A,k}^{\perp}}^TB }
         + \frac{1}{\sigma_{min}(A_k)}
           \FNorm{ \Omega \mathcal{S}U_{A,k}^{\perp}{U_{A,k}^{\perp}}^TB }         \\
\label{eqn:pfCeq1}
   &\leq&  \frac{1}{\sigma_{min}(A_k)}
           \FNorm{U_{A,k}^T\mathcal{S}^T\mathcal{S}U_{A,k}^{\perp}{U_{A,k}^{\perp}}^TB} 
         + \frac{1}{\sigma_{min}(A_k)}
           \TNorm{\Omega} \FNorm{ \mathcal{S}U_{A,k}^{\perp}{U_{A,k}^{\perp}}^TB }  .
\end{eqnarray}
By combining (\ref{eqn:pfCeq1}) with Lemmas~\ref{lemma:matmult1}, 
\ref{lemma:matmult2}, and~\ref{lemma:matmult3}, it follows that
\begin{eqnarray*}
\FNorm{\tilde{X}_{opt} - X_{opt}} 
   &\leq& \sigma_{\min}^{-1}(A_k)\left(\epsilon/2 + 10\epsilon/\sqrt{2} \right) \mathcal{Z}      \\
   &\leq& \frac{8 \epsilon} {\sigma_{\min}(A_k)} \mathcal{Z}     .
\end{eqnarray*}
Equation (\ref{eqn:result3}) follows by setting $\epsilon' = \epsilon/8$ and 
using the value of $r$ assumed by the theorem.

\subsubsection{Proof of Equation (\ref{eqn:result4})}

The error bound provided by (\ref{eqn:result3}) could be quite weak, since 
$\min_{X \in \mathbb{R}^{n \times p}} \FNorm{B - A_kX}$ could be quite close 
or even equal to $\FNorm{B}$, if $B$ has most or all of its ``weight'' outside 
of the column space of $A_k$.
Under a slightly stronger assumption, we will provide a bound 
$\FNorm{\tilde{X}_{opt} - X_{opt}}$ in terms of $\FNorm{X_{opt}}$, thus 
proving (\ref{eqn:result4}).

If we make the additional assumption 
that 
a 
{\em constant fraction} of the ``weight'' of $B$ lies in the subspace spanned 
by the columns of $A_k$, then it follows that
\begin{eqnarray}
                   \mathcal{Z}^2
\nonumber             &=&    \left( \min_{X \in \mathbb{R}^{n \times p}} \FNorm{B - A_kX} \right)^2     \\
\nonumber             &=&    \FNormS{U_{A,k}^{\perp}{U_{A,k}^{\perp}}^TB}          \\
\nonumber             &=&    \FNormS{B} - \FNormS{U_{A,k} U_{A,k}^T B}            \\
\label{eqn:pf4eq1}    &\leq& (\gamma^{-2}-1) \FNormS{U_{A,k} U_{A,k}^T B}           .
\end{eqnarray}
In order to relate $\FNorm{U_{A,k} U_{A,k}^T B}$ and thus $\mathcal{Z}$ to 
$\FNorm{X_{opt}}$ note that 
\begin{eqnarray}
          \FNorm{X_{opt}} 
\nonumber             &=& \FNorm{V_{A,k} \Sigma_{A,k}^{-1} U_{A,k}^T B}                \\
\nonumber             &=&    \FNorm{\Sigma_{A,k}^{-1} U_{A,k}^T B}                 \\
\nonumber             &\geq& \sigma_{\min}(\Sigma_{A,k}^{-1})\FNorm{U_{A,k}^TB}    \\
\label{eqn:pf4eq2}    &=&    \frac{\FNorm{U_{A,k} U_{A,k}^TB}}{{\sigma_{\max}(A_k)}}  .
\end{eqnarray}
By combining (\ref{eqn:result3}) with (\ref{eqn:pf4eq1}) and 
(\ref{eqn:pf4eq2}), we get
\begin{eqnarray*}
\FNorm{\tilde{X}_{opt} - X_{opt}} 
   &\leq& \frac{\epsilon}{\sigma_{\min}(A_k)} \mathcal{Z}                      \\ 
   &\leq& \frac{\epsilon}{\sigma_{\min}(A_k)} \sqrt{\gamma^{-2}-1} \FNorm{U_{A,k} U_{A,k}^T B}     \\
   &\leq& \epsilon \frac{\sigma_{\max}(A_k)}{\sigma_{\min}(A_k)} \sqrt{\gamma^{-2}-1} \FNorm{X_{opt}}    ,
\end{eqnarray*}
which establishes (\ref{eqn:result4}).

\subsubsection{Modifications to the proof with alternate row sampling procedure}
\label{sxn:proof_modifications}

If, in Algorithm~\ref{alg:sample_l2regression}, the rows are sampled with the
\textsc{Expected($c$)} algorithm, then the proof of the claims of
Theorem~\ref{thm:ls_bound} is analogous to the proof described in the four
previous subsections, with the following major exception.
The claims of Lemma~\ref{lemma:matmult1} hold if $r=O(k\log k/\beta\epsilon^2)$ 
rows are chosen with the \textsc{Expected($c$)} algorithm.
To see this, recall that to bound the first claim of 
Lemma~\ref{lemma:matmult1}, we must bound the spectral norm
$\TNorm{U_{A,k}^TU_{A,k} - U_{A,k}^T\mathcal{S}^T\mathcal{S}U_{A,k}}$
in (\ref{eq18}).
If the sampling is performed with the \textsc{Exactly($c$)} algorithm, then
this is bounded in (\ref{eq19}) by the corresponding Frobenius norm, which is 
then bounded with Theorem~\ref{thm:matmult-main-exact}. 
On the other hand, if the sampling is performed with the 
\textsc{Expected($c$)} algorithm, then we can bound (\ref{eq18}) directly with 
the spectral norm bound provided by Theorem~\ref{thm:matmult-main-expected}.

Since the remaining claims of Lemma~\ref{lemma:matmult1} follow from the 
first, they are also valid if $r=O(k\log k/\beta\epsilon^2)$ rows are chosen 
with the \textsc{Expected($c$)} algorithm.
Lemma~\ref{lemma:matmult2} still follows if 
$r=400k/\beta\epsilon^2$, by using the Frobenius norm bound of 
Theorem~\ref{thm:matmult-main-expected}, and
Lemma~\ref{lemma:matmult3} also follows immediately.
The proofs of (\ref{eqn:result2}), (\ref{eqn:result3}), and 
(\ref{eqn:result4}) are identical, and thus Theorem~\ref{thm:ls_bound}, under 
the assumption that the the rows are chosen with the 
\textsc{Expected($c$)} algorithm, follows.

\section{Empirical Evaluation}
\label{sxn:empirical}

Although this is a theoretical paper, it is motivated by applications, and
thus one might wonder about the empirical applicability of our methods.
For example, if we want to do as well as the best rank $k=100$
(respectively, $k=10$) approximation, with relative error bound
$\epsilon = 0.1$, then our main theorem samples $3.2$ billion
(respectively $32$ million) columns of the matrix $A$ using the
\textsc{Exactly}$(c)$ algorithm. Of course, our main theorem states
that it in order to obtain our strong provable worst-case
relative-error guarantees it \emph{suffices} to choose that many
columns. But, it would be a source of concern if anything like that
number of columns is needed in ``real'' scientific and internet data
applications.

In this section, we provide an empirical evaluation of the
performance of our two main sampling procedures both for CX and CUR
decompositions. In particular, we will evaluate how well the
proposed column/row selection strategies perform at capturing the
Frobenius norm for matrices derived from DNA Single Nucleotide
Polymorphism (SNP) analysis, recommendation system analysis, and
term-document analysis. By applying our algorithms to data sets
drawn from these three diverse domains of modern data analysis, we
will demonstrate that we can obtain very good Frobenius norm
reconstruction by sampling a number of columns and/or rows that
equals a small constant, e.g., $2$ or $3$ or $4$ (as opposed to,
e.g., a million or a billion), times the rank parameter $k$.

\subsection{Details of Our Empirical Evaluation}

The empirical evaluation of our CX and CUR matrix decompositions has
been performed using the following two types of column/row selection methods:
\begin{itemize}
\item
``Subspace sampling'' (with replacement) using the \textsc{Exactly}$(c)$
algorithm; and
\item
``Subspace sampling'' (without replacement) using the \textsc{Expected}$(c)$
algorithm.
\end{itemize}
In addition, the empirical evaluation has been performed on the following
three data sets:
\begin{itemize}
\item
Matrices derived from the DNA SNP HapMap data~\cite{IHMC05,Paschou07a}--see
Section~\ref{sxn:empirical:hapmap};
\item
A matrix derived from the Jester recommendation system
corpus~\cite{GRGP01,MMD06}--see Section~\ref{sxn:empirical:jester}; and
\item
A matrix derived from the Reuters term-document
corpus~\cite{LYRL04,DDHJM07}--see Section~\ref{sxn:empirical:reuters}.
\end{itemize}
We have chosen these three data sets on which to evaluate the empirical
applicability of our algorithms for three reasons:
first, these three application domains are representative of a wide range of
areas of modern scientific and internet data analysis;
second, these matrices are all approximately (to a greater or lesser extent)
low-rank, and they are all data for which spectral methods such as low-rank
approximations have been successfully applied;
and third, we have already (with collaboraters from these application
areas) applied our algorithms to these data
sets~\cite{MMD06,Paschou07a,DDHJM07}.
In these data application papers~\cite{MMD06,Paschou07a,DDHJM07}, we
have shown that our main CX and CUR decomposition algorithms (either the
algorithms for which we have provable performance guarantees and/or greedy
variants of these basic algorithms) perform well on tasks such as
classification, denoising, reconstruction, prediction, and clustering---tasks
that are of more immediate interest to data practitioners than simply
capturing the norm of the data matrix.

In this section, however, we we will restrict ourselves to an empirical
evaluation of our two main theorems.
To do so, we will fix a rank parameter $k$, and we will present plots of the
Frobenius norm error (normalized by $||A-A_k||_F$), as a function of the
number of samples chosen.
For example, we will consider $\Theta_1 \equiv ||A-CC^+A||_F/||A-A_k||_F$,
where $A_k$ is the best rank-$k$ approximation to the matrix $A$, as a
function of the number $c$ of columns chosen.
This ratio corresponds to the quantity that is bounded by $1+\epsilon$ in
Theorem~\ref{thm:relerr_CX}.
For $c=k$, this quantity will be no less than $1$; of course, if we choose
$c > k$ columns then this ratio may be less than $1$.
Following the remark after Theorem~\ref{thm:relerr_CX}, we will also consider
$\Theta_2 \equiv ||A-CC^+A_k||_F/||A-A_k||_F$.
This ensures that the approximation has rank no greater than $k$ (which is of
interest in certain applications), and thus the plotted ratio will clearly be
no less than $1$, for every value of $c$.
We will also consider $\Theta_3 \equiv ||A-CUR||_F/||A-A_k||_F$, which
corresponds to the quantity that is bounded by $1+\epsilon$ in
Theorem~\ref{thm:relerr_CUR}.

Two technical points should be noted about these plots in the upcoming
subsections.
First, we ran our CX or CUR decomposition algorithm several---e.g., three or
five, depending on the size of the data being plotted---times (corresponding,
say, to multiple runs to boost the $\delta$ failure probability) and the
minimum value over these repetitions was returned; this was repeated several times and the average of
those values is plotted.
Second, for the plots of $||A-CUR||_F/||A-A_k||_F$, the number of rows
selected is set to be twice the corresponding number of columns selected;
optimizing over this would lead to marginally better performance than that
presented.

\subsection{DNA SNP HapMap Data}
\label{sxn:empirical:hapmap}

Our first dataset comes from the field of human genetics.
The HapMap project, a continuation of the Human Genome project, aims to map
the loci in the human genome that differ between individuals~\cite{IHMC05}.
The HapMap project focuses on the so-called SNPs (Single Nucleotide
Polymorphisms), which are a very common type of variation in the genome
(nearly $10^7$ such loci have been identified in the human genome).
Significant motivation exists in the genetics community for minimizing the
number of SNPs that must be assayed, and in~\cite{Paschou07a}, we demonstrated
how CUR-type methods may be used to efficiently reconstruct unassayed SNPs
from a small number of assayed SNPs.

Both in~\cite{Paschou07a} and here, we consider two regions of the genome
known as HOXB and 17q25.
Three populations were studied for each region: Yoruban, a sub-Saharan African
population; a European population; and a joint Japanese/Chinese population.
Each population had $90$ individuals, each corresponding to a row of
the input matrix.
Columns of each matrix correspond to SNPs within the HOXB or 17q25 regions.
The genotypic data were encoded appropriately in order to be converted to
numeric data in the form of matrices.
(Careful preprocessing was done to remove fixed SNPs, as well as SNPs with
too many missing entries, etc.)
The HapMap project provided data on $370$ SNPs in 17q25 and $571$ SNP in
HOXB~\cite{IHMC05}.
Thus, for example, our matrix for the Yoruban population in HOXB is a
$90 \times 571$ matrix, whose entries are in the set $\{-1,0,+1\}$.%
\footnote{The encoding should be interpreted as follows: each SNP consists of
two alleles (nucleotide bases); these bases are the same for all humans. Say
that these bases are A and G.  Then a value of $+1$ corresponds to an
individual whose genotype (pair of alleles) is AA, a value of $0$ corresponds
to an individual whose genotype is AG or GA, and a value of $-1$ corresponds
to an individual whose genotype is GG.}
The other data matrices are of similar (not extremely large) size.
See~\cite{Paschou07a} and references therein for details.

Data not presented indicate that for all three populations and for
both genomic regions, the data possess a great deal of linear
structure. For example, in the 17q25 matrices, one needs $9$, $9$,
and $7$ singular vectors to capture $80\%$ of the Frobenius norm for
the Yoruban, European, and the Japanese/Chinese populations,
respectively; and one needs $18$, $16$, and $13$ singular vectors,
respectively, to capture $90\%$. The matrices for the HOXB region of
the genome are even more redundant; one needs only $7$, $6$, and $4$
singular vectors, respectively, to capture $80\%$ of the Frobenius
norm.

\begin{figure}[t]
\begin{center}
\includegraphics[height=4.5in]{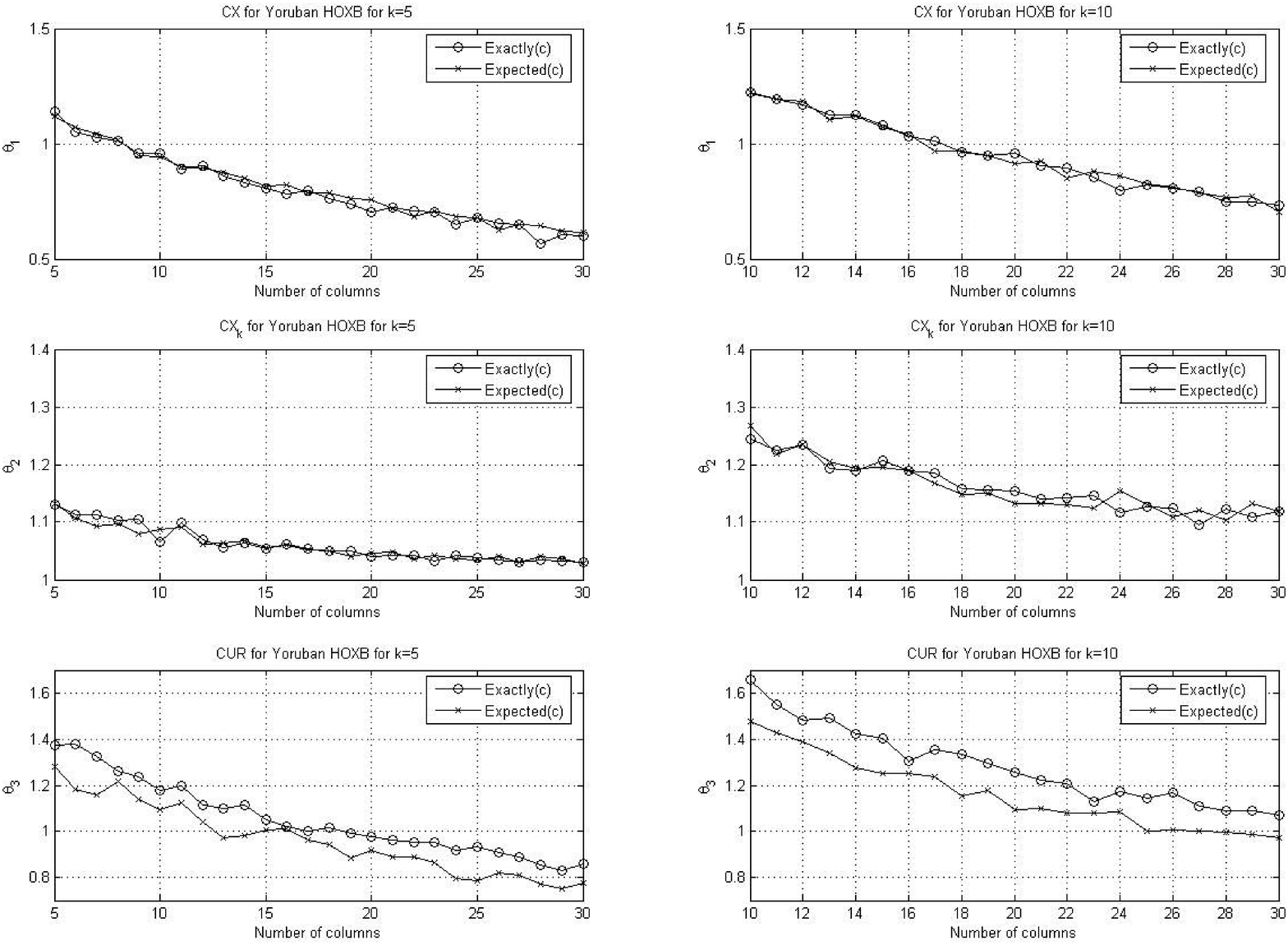}
\end{center}
\caption{Reconstruction error for the Yoruban population in the HOXB
region of the genome.  Shown are $\Theta_1$, $\Theta_2$, and
$\Theta_3$ (as defined in the text) for two values of the rank
parameter $k$.  The X-axis corresponds to the number of columns
sampled with the \textsc{Exactly}$(c)$ algorithm or the
\textsc{Expected}$(c)$ algorithm.} \label{fig:snp1}
\end{figure}

In Figure~\ref{fig:snp1}, data are presented for the Yoruban HOXB
data matrix. Each of the six subfigures presents a plot of the
Frobenius norm error as a function of the number $c$ of samples
chosen. In particular, for two values of the rank parameter, i.e.,
$k=5$ and $k=10$, the ratio $\Theta_i =
||A-A^{\prime}||_F/||A-A_k||_F$ is plotted, where:
$A^{\prime}=CC^+A$ for $i=1$; $A^{\prime}=CC^+A_k$ for $i=2$; and
$A^{\prime}=CUR$ for $i=3$. Clearly, in all these cases, only modest
oversampling is needed to capture ``nearly all'' of the dominant
part of the spectrum of the data matrix. For example, for $k=5$: if
$c=5$ then $\Theta_1=1.12$; if $c\ge6$ then $\Theta_1<1.1$; and if
$c \ge 9$ then $\Theta_1<1.0$. Similarly, for $k=10$: if $c=10$ then
$\Theta_1=1.22$; if $c\ge 15$ then $\Theta_1<1.1$; and if $c\ge18$
then $\Theta_1<1.0$. Similar results hold if the projection onto the
span of the columns is regularized through a rank-$k$ space and also
if rows are chosen after the columns. For example, for $k=10$: if
$c\ge16$ then $\Theta_2<1.2$ and if $c \gtrsim 30$ then
$\Theta_2<1.1$. Similarly, even though the computations for
$\Theta_3$ are slightly worse and somewhat noisier due to the second
level of sampling (columns and then rows), the results still show
that only modest oversampling (of $c$ relative to $k$) is needed.
For example, if $k=10$, then $\Theta_3<1.1$ if $c \ge 20$ or $c \ge
28$, depending on precisely how the columns are chosen.
Interestingly, in this last case, not only are the plots noisier,
but the \textsc{Expected}$(c)$ algorithm and the
\textsc{Exactly}$(c)$ algorithm seem to lead to (slightly) different
results as a function of $c$.

\begin{figure}[t]
\begin{center}
\includegraphics[height=4.5in]{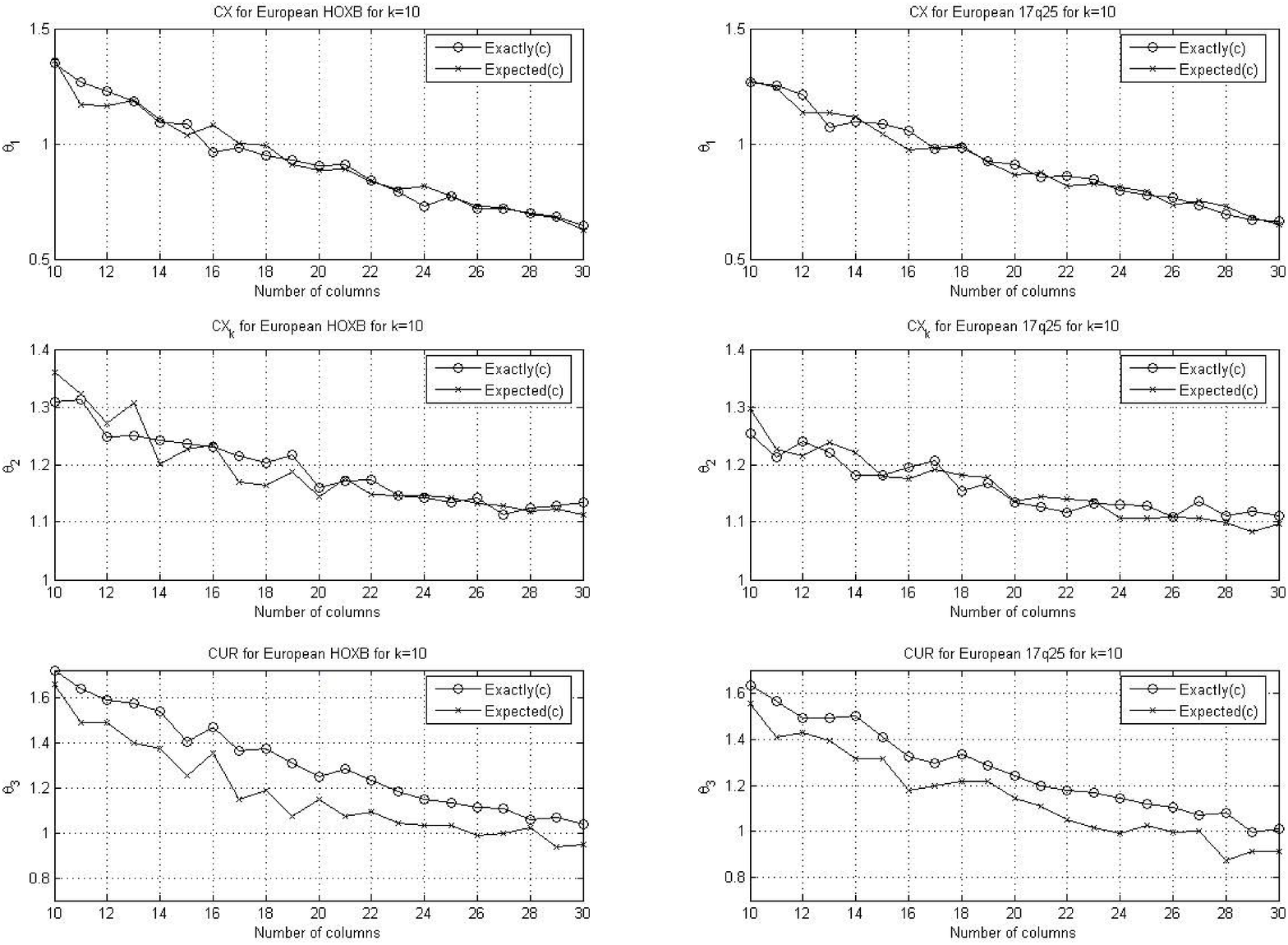}
\end{center}
\caption{Reconstruction error for the European population in both
the HOXB and 17q25 regions of the genome.  Shown are $\Theta_1$,
$\Theta_2$, and $\Theta_3$ (as defined in the text) for $k=10$.  The
X-axis corresponds to the number of columns sampled with the
\textsc{Exactly}$(c)$ algorithm or the \textsc{Expected}$(c)$
algorithm.} \label{fig:snp2}
\end{figure}

Qualitatively similar results are seen for the other populations and
the other genomic regions. For example, in Figure~\ref{fig:snp2},
data are presented for the European population for both the HOXB and
the 17q25 regions of the genome for the value of the rank parameter
$k=10$. For the HOXB region, $\Theta_1=1.36$ if $c=10$ (this is
higher than for the corresponding Yoruban data), $\Theta_1 < 1.0$ if
$c \gtrsim 17$ (this is similar to the corresponding Yoruban data),
and $\Theta_1=0.62$ if $c=30$ (this is less than the Yoruban data).
Similar trends are seen for $\Theta_2$ and $\Theta_3$ and also for
the 17q25 region. In all cases, only very modest oversampling is
needed for accurate Frobenius norm reconstruction. Data not
presented indicate that the data for the joint Japanese/Chinese
population is quite similar or slightly better to those results
presented.

\subsection{Recommendation System Jester Data}
\label{sxn:empirical:jester}

Our second dataset comes from the field of recommendation system analysis, in
which one is typically interested in making purchase recommendations to a user
at an electronic commerce web site~\cite{GRGP01}.
Collaborative methods (as opposed to content-based or hybrid) involve
recommending to the user items that people with similar tastes or preferences
liked in the past.
Many collaborative filtering algorithms represent a user as an $n$
dimensional vector, where $n$ is the number of distinct products, and where the
components of the vector are a measure of the rating provided by that user for
that product.
Thus, for a set of $m$ users, the user-product ratings matrix is an
$m \times n$ matrix $A$, where $A_{ij}$ is the rating by user $i$ for product
$j$ (or is null if the rating is not provided).

The so-called Jester joke dataset is a commonly-used benchmark for
recommendation system research and development~\cite{GRGP01}.
In~\cite{MMD06}, we applied a CUR decomposition on this data to the problem
of reconstructing missing entries and making accurate recommendations.
Here, we consider the $m=14,116$ (out of ca.  $73,000$) users who rated all of
the $n=100$ products (i.e., jokes) in the Jester data.
The entries in this $14,116 \times 100$ matrix $A$ are real numbers between
$-10$ and $+10$ that represent the user's rating of a product.

\begin{figure}[t]
\begin{center}
\includegraphics[height=4.5in]{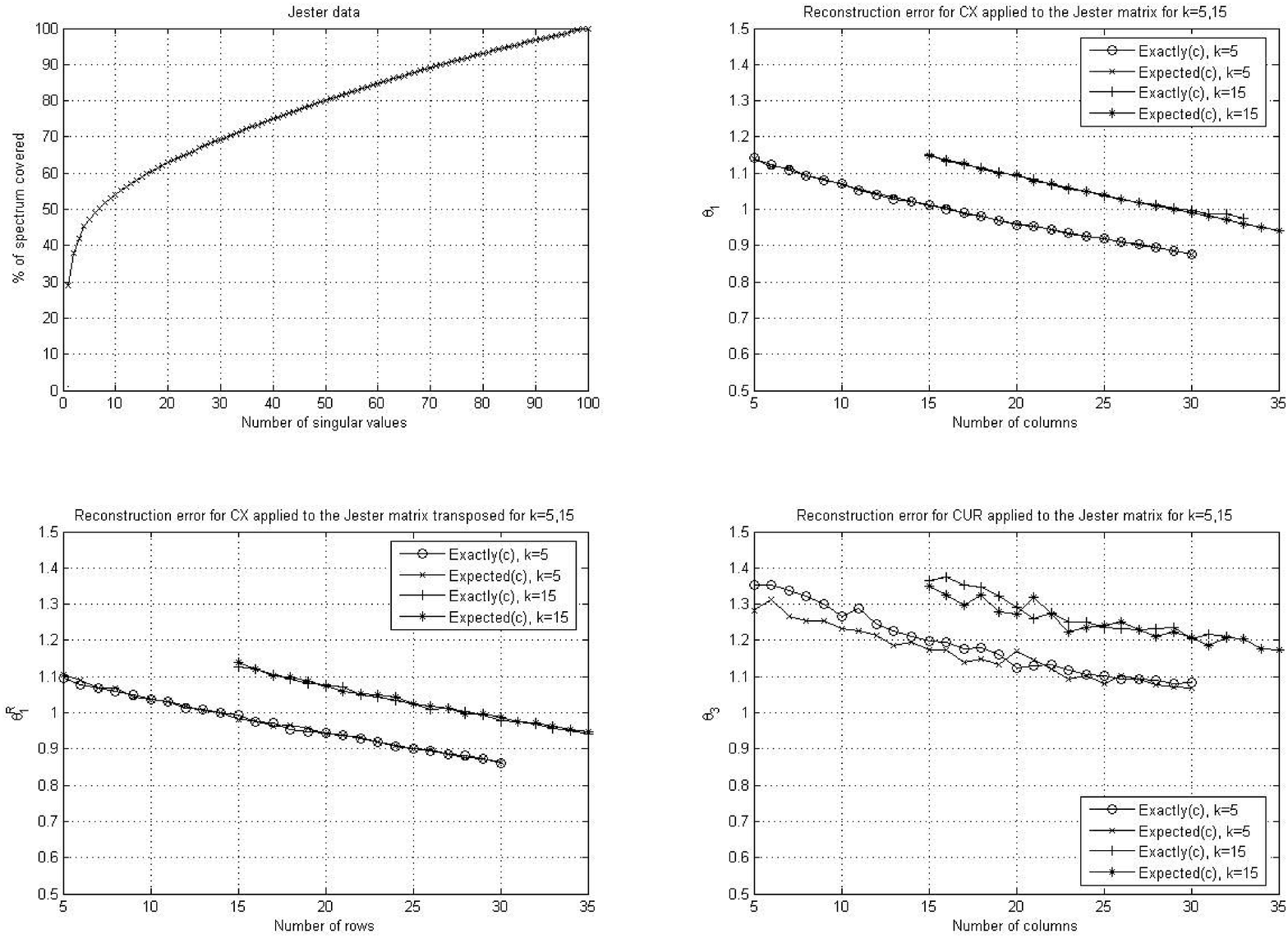}
\end{center}
\caption{Empirical results for the Jester recommendation system data.  Shown
are: the percentage of the Frobenius norm captured as a function of the
number of singular components; $\Theta_1$ for sampling columns from $A$ for
$k=5$ and $k=15$; the analogue of $\Theta_1$ for selecting rows from $A$
(i.e., $\Theta_1$ for sampling columns from $A^T$); and $\Theta_3$ for
selecting columns and then rows for $k=5$ and $k=15$.}
\label{fig:jester}
\end{figure}

Figure~\ref{fig:jester} presents the empirical results for the
Jester recommendation system data. The rank of the $14,116 \times
100$ matrix is $100$, and although only $7$ singular vectore are
needed to capture $50\%$ of the Frobenius norm, $50$ are needed to
capture $80\%$, and $73$ are needed to capture $90\%$. Thus, the
spectrum and shape of this matrix (this matrix is very rectangular)
are very different from that of the matrices of the previous
subsection.

Figure~\ref{fig:jester} presents reconstruction error results for
selecting columns (i.e., products or jokes), for selecting rows
(i.e., users), and for selecting both columns and rows
simultaneously. For example, when selecting columns from $A$, if
$k=15$, then $\Theta_1=1.14$ if $c=15$, $\Theta_1 \le 1$ if $c
\gtrsim 29$, and $\Theta_1=0.99$ when $c=30$. Although the matrix is
very rectangular, quantitatively very similar results are obtained
for the analogue of $\Theta_1$ (called $\Theta_1^R$ in the figure)
if rows are sampled (or, equivalently if columns are sampled from
$A^T$). Thus, when our main CX decomposition is applied to either
$A$ or to $A^T$, a small number of columns (products) or rows
(users), capture most of the Frobenius norm of $A$ that is captured
by the best rank $k$ approximation to $A$. A similar result holds
for the simultaneously choosing columns and rows of $A$ (both users
and products), and applying our CUR approximation algorithm. As with
the data of the previous subsection, the data for $\Theta_3$ are
much noisier when both columns and rows are chosen, but even in this
case $\Theta_3\le 1.1$ for $k=5$ if $c\gtrsim25$ and $\Theta_3\le
1.2$ for $k=15$ if $c\gtrsim30$. In all these cases, data not
presented indicate that qualitatively similar (but shifted) results
are obtained for higher values of the rank parameter $k$.

\subsection{Term-Document Reuters Data}
\label{sxn:empirical:reuters}

Our third data set comes from the field of text categorization and
information retrieval. In these applications, documents are often
represented as a so-called ``bag of words'' and a vector space model
is used. In 2000, Reuters Ltd made available a large collection of
Reuters News stories for use in research and development of natural
language processing, information retrieval, and machine learning
systems. This corpus, known as "Reuters Corpus, Volume 1" or RCV1,
is significantly larger (it contains over $800,000$ news items from
1996-97) than the older, well-known Reuters-21578 collection which
has been heavily used in the text classification
community~\cite{LYRL04}. In~\cite{DDHJM07}, we considered the
problem of feature selection for improved classification, and we
compared a CX-like column selection procedure to several traditional
methods. The data come with class labels and possess a hierarchical
class structure (which we used in~\cite{DDHJM07} but) which we
ignored here. Here, we used the {\em ltc}-normalized term-document
matrix and the training data from the one test-train split provided
by Lewis~{\it et al.}~\cite{LYRL04}. Thus, the Reuters matrix we
considered here is a (very sparse) $47,236 \times 23,149$ matrix
whose elements are real numbers between $0$ and $1$ that represent a
normalized frequency.

\begin{figure}[t]
\begin{center}
\includegraphics[height=4.5in]{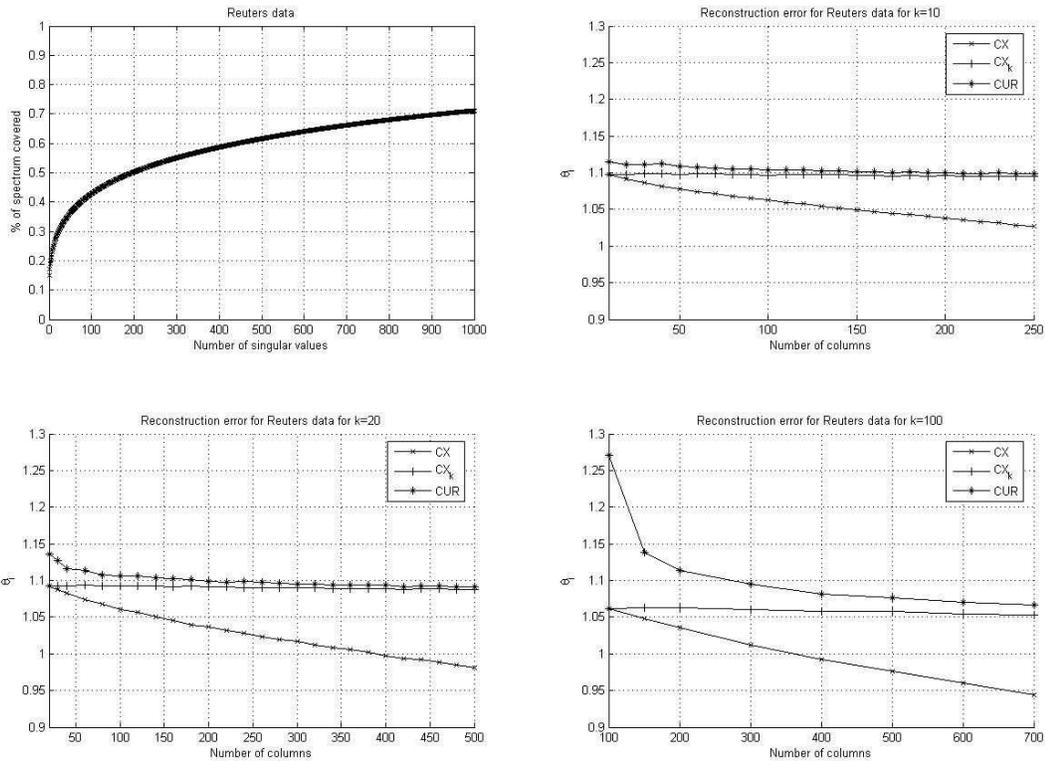}
\end{center}
\caption{Empirical results for the Reuters term-document data.  Shown are: the
percentage of the Frobenius norm captured as a function of the number of
singular components; $\Theta_1$, $\Theta_2$, and $\Theta_3$ as a function of
the number of sampled columns and/or rows for three different values of the
rank parameter $k$.}
\label{fig:reuters}
\end{figure}

Figure~\ref{fig:reuters} presents the empirical results for the
Reuters term-document data. Note that this data is not only much
larger than the data from the previous two subsections, it is also
less well approximated by a low rank matrix. Less than $50\%$ of the
Frobenius norm is captured by the first $k=100$ singular components,
and less than $80\%$ is captured by the first $k=1500$ singular
components. (Nevertheless, spectral methods have frequently been
applied to this data.) The matrix is very sparse, and performing
computations is expensive in terms of space and time (due to
multiple randomized trials and since the dense matrices of singular
vectors are large) if the rank parameter $k$ is chosen to be more
than a few hundred. Thus, to demonstrate the empirical applicability
of our main algorithms, we considered several smaller values of $k$.
Here, we report results for: $k=10$ and $c=10$ to $250$; for $k=20$
and $c=20$ to $500$; and for $k=100$ and $c=100$ to $700$. Note that
we report results only for columns chosen with the
\textsc{Expected}$(c)$ algorithm; initial unreported computations on
several smaller systems indicate that very similar results will be
obtained with the \textsc{Exactly}$(c)$ algorithm.

In all of these cases, and for all values of $\Theta_1$, $\Theta_2$,
and $\Theta_3$, only modest oversampling leads to fairly small
reconstruction error. The worst data point reported was for
$\Theta_3=1.272$ for $k=100$ and $c=100$, and even in that case
$\Theta_3 < 1.1$ for $c\ge300$. Interestingly, all the curves tend
to decrease somewhat more slowly (as a function of oversampling $c$,
relative to $k$) than the corresponding curves in the previous
subsections do. Note that $\Theta_1$ does not decrease below $1.0$
for $k=10$ until after $c=500$; for $k=20$, it drops below $1.0$ by
$c=400$, and for $k=100$ (which obviously captures the largest
fraction of the Frobenius norm) it drops below $1.0$ at $c \approx
350$. Thus, this phenomenon is likely related to the degree to which
the chosen value for the rank parameter $k$ captures a reasonable
fraction of the norm of the original matrix. Nevertheless, in all
cases, we can achieve $\Theta_i <1.1$ with only a modest degree of
oversampling $c$ relative to $k$.

\section{Conclusion}
\label{sxn:conclusion}

We have presented and analyzed randomized algorithms for computing low-rank 
matrix approximations that are explicitly expressed in terms of a small number 
of columns and/or rows of the input matrix.
These algorithm achieve relative-error guarantees, whereas previous algorithms 
for these problems achieved only additive-error guarantees.
These algorithms randomly sample in a novel manner we call ``subspace 
sampling,'' and their analysis amounts to approximating a generalized $\ell_2$ 
regression problem by random sampling.
As described in Section~\ref{sxn:intro:data_apps} and
in~\cite{Paschou07a,MMD06}, such low-rank matrix approximations have 
numerous applications for the improved analysis of data.

We conclude with several open problems.
\begin{itemize}
\item 
To what extent do the results of this paper generalize to other matrix norms?
\item 
What hardness results can be established for the optimal choice of columns 
and/or rows?
\item 
Does there exist a deterministic approximation algorithm for
either of the problems we consider?
\item
Does there exist an efficient deterministic algorithm to choose columns and/or 
rows that exactly or approximately optimize the maximum volume of the induced 
parallelepiped?
(As pointed out to us by an anonymous reviewer, \cite{GE96,Tyr04b} provide 
such procedures; it would be interesting to see if bounds of the form we 
prove can be established for the algorithms of \cite{GE96,Tyr04b}).
\item
Can we formulate a simple condition that we can check after we have sampled 
the columns and/or rows to determine whether we have achieved a $1+\epsilon$ 
approximation with that sample?
\item
Can we obtain similar algorithms and comparable bounds for formulations of 
these problems that include regularization and/or conditioning?
\item
What heuristic variants of these algorithms are most appropriate in different 
application domains?
\item
Are the algorithms presented in this paper numerically stable?
\end{itemize}

\vspace{0.02in}

\noindent
\textbf{Acknowledgments:}
We would like to thank: 
Sariel Har-Peled for writing up his results amidst travel in India~\cite{HarPeled06_relerr_DRAFT};
Amit Deshpande and Santosh Vempala for graciously providing a copy of~\cite{DV06_relerr_TR}; 
and two anonymous reviewers for useful comments.


\appendix
\section{Approximating Matrix Multiplication}
\label{sxn:matrix_multiply}

In this section, we describe two complementary procedures for randomly 
sampling (and rescaling) columns and/or rows from an input matrix.
Then, we describe an algorithm for approximating the product of two matrices
by randomly sampling columns and rows from the input matrices using one of 
the two sampling procedures.

\subsection{Sampling Columns and Rows from Matrices}
\label{sxn:matrix_multiply:makeSD}

We describe two simple algorithms for randomly sampling a set of columns from 
an input matrix.  
Each algorithm takes as input an $m \times n$ matrix $A$ and a probability 
distribution $\{p_i\}_{i=1}^{n}$,
and each constructs a matrix $C$ consisting of a rescaled copy of small number 
of columns from $A$.
Clearly, each algorithm can be modified to sample rows from a matrix.
The first algorithm is the \textsc{Exactly($c$)} algorithm, which is described 
in Algorithm~\ref{alg:SDconstruct_exact} using the sampling matrix formalism 
described in Section~\ref{sxn:review_la}.
In this algorithm, $c$ columns \emph{exactly} of $A$ are chosen in $c$ i.i.d. 
trials, where in each trial the $i$-th column of $A$ is picked with 
probability $p_i$.
Note that because the sampling is performed with replacement a single column 
of $A$ may be included in $C$ more than once.
The second algorithm is the \textsc{Expected($c$)} algorithm, which is 
described in Algorithm~\ref{alg:SDconstruct_expected}, also using the sampling 
matrix formalism described in Section~\ref{sxn:review_la}.
In this algorithm, at most $c$ columns 
\emph{in expectation} of $A$ are chosen by including the $i$-th column of $A$ 
in $C$ with probability $\tilde{p}_i = \min \{1,cp_i\}$.
Note that the exact value of the number of columns returned is not known 
before the execution of this second algorithm; we do not perform an analysis 
of this random variable.

\begin{algorithm}
\begin{framed}

\SetLine

\AlgData{
$A \in \mathbb{R}^{m \times n}$,
$p_i \geq 0, i\in[n]$ s.t. $\sum_{i \in [n]}p_i=1$, 
positive integer $c \leq n$.
}

\AlgResult{
Sampling matrix $S$, 
rescaling matrix $D$, and 
sampled and rescaled columns $C$.
}

Initialize $S$ and $D$ to the all zeros matrices.

\For{$t=1,\ldots,c$}{
   Pick $i_t \in [n]$, where $\Prob(i_t = i) = p_i$\;
   $S_{i_t t} = 1$\;
   $D_{tt} = 1/\sqrt{cp_{i_t}}$\; 
}

$C = ASD$\;

\end{framed}
\caption{
The \textsc{Exactly($c$)} algorithm to create $S$, $D$, and $C$.
}
\label{alg:SDconstruct_exact}
\end{algorithm}
\begin{algorithm}
\begin{framed}

\SetLine

\AlgData{
$A \in \mathbb{R}^{m \times n}$,
$p_i \geq 0, i\in[n]$ s.t. $\sum_{i \in [n]}p_i=1$,
positive integer $c \leq n$.
}

\AlgResult{
Sampling matrix $S$, 
rescaling matrix $D$, and
sampled and rescaled columns $C$.
}

Initialize $S$ and $D$ to the all zeros matrices.

$t=1$\;
\For{$j=1,\ldots,n$}{
   Pick $j$ with probability $\min\{1,cp_j\}$\;
   \If{$j$ is picked} {
      $S_{jt} = 1$\;
      $D_{tt} = 1/\min\{1,\sqrt{cp_{j}}\}$\;
      $t = t + 1$\;
   }
}

$C = ASD$\;

\end{framed}
\caption{
The \textsc{Expected($c$)} algorithm to create $S$, $D$, and $C$.
}
\label{alg:SDconstruct_expected}
\end{algorithm}

\subsection{Approximate Matrix Multiplication Algorithms}
\label{sxn:matrix_multiply:alg}

Algorithm~\ref{alg:matrix_multiply} takes as input two matrices $A$ and $B$, 
a number $c \le n$, and a probability distribution $\{p_i\}_{i=1}^{n}$ over 
$[n]$.  
It returns as output two matrices $C$ and $R$, where the columns of $C$ are a 
small number of sampled and rescaled columns of $A$ and where the rows of $R$ 
are a small number of sampled and rescaled rows of $B$.
The sampling and rescaling are performed by calling either the 
\textsc{Exactly($c$)} algorithm or the \textsc{Expected($c$)} algorithm. 
When the \textsc{Exactly($c$)} algorithm is used to choose column-row pairs in 
Algorithm~\ref{alg:matrix_multiply}, this is identical to the algorithm 
of \cite{dkm_matrix1}.
In particular, note that \emph{exactly} $c$ column-row pairs are chosen, and 
a column-row pair could be included in the sample more than once.
When the \textsc{Expected($c$)} algorithm is used to choose column-row pairs 
in Algorithm~\ref{alg:matrix_multiply} this is a minor variation of the 
algorithm of \cite{dkm_matrix1}.
In particular, the main difference is that at most $c$ column-row pairs 
\emph{in expectation} are chosen, and no column-row pair is included in the 
sample more than once.

\begin{algorithm}[h]

\begin{framed}

\SetLine

\AlgData{
$A \in \mathbb{R}^{m \times n}$, 
$B \in \mathbb{R}^{n \times p}$, 
$\{p_i\}_{i=1}^n$ such that $\sum_{i=1}^n p_i = 1$, 
$c \leq n$.
}

\AlgResult{
$C \in \mathbb{R}^{m \times c}$, 
$R \in \mathbb{R}^{c \times p}$.
}

\begin{itemize}
\item
Form the matrix $C=ASD$ by sampling according to
$\{p_i\}_{i=1}^{n}$ with the \textsc{Exactly($c$)} algorithm 
or with the \textsc{Expected($c$)} algorithm\;
\item
Form the matrix $R =DS^TB$ from the corresponding rows of $B$\;
\end{itemize}

\end{framed}

\caption{
A fast Monte-Carlo algorithm for approximate matrix multiplication.
}

\label{alg:matrix_multiply}

\end{algorithm}

The next two theorems are our basic quality-of-approximation results for 
Algorithm~\ref{alg:matrix_multiply}.  
Each states that, under appropriate assumptions, 
$ CR = A SD DS^T B \approx AB $.
The most interesting of these assumptions is that the sampling probabilities 
used to randomly sample the columns of $A$ and the corresponding rows of $B$ 
are nonuniform and depend on the product of the Euclidean norms of the columns 
of $A$ and/or the corresponding rows of $B$.
For example, consider sampling probabilities $\left\{ p_i \right\}_{i=1}^{n}$ 
such that 
\begin{equation}
\label{eqn:nearly_optimal_probs}
p_i \geq \beta \frac{             \VTTNorm{A^{(i)}}\VTTNorm{B_{(i)}}}
                    {\sum_{j=1}^n \VTTNorm{A^{(j)}}\VTTNorm{B_{(j)}}}   ,     
\end{equation}
for some $\beta \in (0,1]$.
Sampling probabilities of the form (\ref{eqn:nearly_optimal_probs}) use 
information from the matrices $A$ and $B$ in a very particular manner.
If $\beta=1$, they are optimal for approximating $AB$ by $CR$ in a sense made 
precise in \cite{dkm_matrix1}.
Alternatively, sampling probabilities $\left\{ p_i \right\}_{i=1}^{n}$ such 
that 
\begin{equation}
\label{eqn:nonoptimal_probs}
p_i \geq \beta \frac{             \VTTNormS{A^{(i)}}}{\FNormS{A}}   ,
\end{equation}
for some $\beta \in (0,1]$, are also of interest in approximating the product 
$AB$ by $CR$ if, e.g., only information about $A$ is easily available.

The following theorem is our main quality-of-approximation result for 
approximating the product of two matrices with 
Algorithm~\ref{alg:matrix_multiply}, 
when column-row pairs are sampled using the \textsc{Exactly($c$)} algorithm.
Its proof (and the statement and proof of similar stronger results) may be 
found in \cite{dkm_matrix1}.

\begin{theorem}
\label{thm:matmult-main-exact} 
Suppose $A \in \mathbb{R}^{m \times n}$, $B \in \mathbb{R}^{n \times p}$, and
$c \le n$.
Construct $C$ and $R$ with Algorithm~\ref{alg:matrix_multiply}, 
using the \textsc{Exactly($c$)} algorithm.
If the sampling probabilities $\left\{ p_i \right\}_{i=1}^{n}$ used by the 
algorithm are of the form (\ref{eqn:nearly_optimal_probs}) or 
(\ref{eqn:nonoptimal_probs}), then
$$
\Expect{\FNorm{AB-CR}} \leq \frac{1}{\sqrt{\beta c}}\FNorm{A}\FNorm{B}  .
$$
\end{theorem}

The following theorem is our main quality-of-approximation result for 
approximating the product of two matrices with 
Algorithm~\ref{alg:matrix_multiply}, 
when column-row pairs are sampled using the \textsc{Expected($c$)} algorithm.
The Frobenius norm bound (\ref{eqn1:thm:matmult-main-expected}) is new, and 
the spectral norm bound (\ref{eqn2:thm:matmult-main-expected}) is due to 
Rudelson and Vershynin, who proved a similar result in a more general 
setting~\cite{Rud99,V03,RV06_DRAFT}.

\begin{theorem}
\label{thm:matmult-main-expected} 
Suppose $A \in \mathbb{R}^{m \times n}$, $B \in \mathbb{R}^{n \times p}$, and
$c \le n$. 
Construct $C$ and $R$ with Algorithm~\ref{alg:matrix_multiply}, 
using the \textsc{Expected($c$)} algorithm.
If the sampling probabilities $\left\{ p_i \right\}_{i=1}^{n}$ used by the 
algorithm are of the form (\ref{eqn:nearly_optimal_probs}) or 
(\ref{eqn:nonoptimal_probs}), then
\begin{equation}
   \label{eqn1:thm:matmult-main-expected} 
   \Expect{\FNorm{AB-CR}} 
      \leq \frac{1}{\sqrt{\beta c}}\FNorm{A}\FNorm{B}  .
\end{equation}
If, in addition, $B=A^T$, then
\begin{equation}
   \label{eqn2:thm:matmult-main-expected} 
   \Expect{ \TNorm{AA^T - CC^T} }
      \leq O(1)\sqrt{\frac{\log c}{\beta c}} \FNorm{A} \TNorm{A}   .
\end{equation}
\end{theorem}
\begin{Proof}
Equation (\ref{eqn2:thm:matmult-main-expected}) follows from the analysis of 
Rudelson and Vershynin, who considered spectral norm bounds on approximating 
the product of two matrices \cite{Rud99,V03,RV06_DRAFT}.
Note that they considered approximating the product $AA^T$ by sampling with 
respect to probabilities of the form (\ref{eqn:nonoptimal_probs}) with 
$\beta = 1$, but the analysis for general $\beta \in (0,1]$ is analogous.

Next, we prove that for any set of probabilities $\{p_i\}_{i=1}^{n}$ the 
following holds:
\begin{equation}
\label{eqn1:prf:thm:matmult-main-expected}
\Expect{\FNormS{AB - CR}} 
   \leq \frac{1}{c} 
        \sum_{j=1}^n \frac{\VTTNormS{A^{(j)}} \VTTNormS{B_{(j)}}}{p_j}.
\end{equation}
Equation (\ref{eqn1:thm:matmult-main-expected}) follows from 
(\ref{eqn1:prf:thm:matmult-main-expected}) by using Jensen's inequality and
using the form of the sampling probabilities (\ref{eqn:nearly_optimal_probs})
and (\ref{eqn:nonoptimal_probs}).

To establish (\ref{eqn1:prf:thm:matmult-main-expected}), recall that the 
sampling is performed with the \textsc{Expected($c$)} algorithm.
Let $I_j$, $j \in [n]$ be the indicator variable that is set to $1$ if the 
$j$-th column of $A$ and the $j$-th row of $B$ are sampled (with probability 
$\min\{1,cp_j\}$) and is set to $0$ otherwise. 
Recall that if $I_j=1$, we scale both the $j$-th column of $A$ and the $j$-th 
row of $B$ by $1/\sqrt{\min\{1,cp_j\}}$.
Thus,
\begin{eqnarray}
\label{eqn:lab1} 
\FNormS{AB - CR} 
   = \FNormS{AB - ASDDS^TB} 
   = \FNormS{\sum_{j=1}^n \left(1-\frac{I_j}{\min\{1,cp_j\}}\right) A^{(j)}B_{(j)}}.
\end{eqnarray}
Clearly, if $\min\{1,cp_j\}=1$, then $I_j = 1$ with probability 1, and 
$1-I_j/\min\{1,cp_j\}=0$. Thus, we can focus on the set of indices 
$\Lambda = \{j \in [n]: cp_j < 1\} \subseteq [n]$.
By taking the expectation of both sides of (\ref{eqn:lab1}), it follows that
\begin{eqnarray}
\Expect{\FNormS{AB - CR}} 
\nonumber &=& \Expect{\FNormS{\sum_{j \in \Lambda} 
                      \left(1-\frac{I_j}{cp_j}\right) A^{(j)}B_{(j)}}}   \\
\nonumber &=& \Expect{\sum_{i_1=1}^m \sum_{i_2=1}^p \left(\sum_{j \in \Lambda}
                      \left(1-\frac{I_j}{cp_j}\right) 
                      A^{(j)}B_{(j)}\right)_{i_1 i_2}^2 }   \\
\nonumber &=& \Expect{\sum_{i_1=1}^m \sum_{i_2=1}^p \left(\sum_{j \in \Lambda}
                      \left(1-\frac{I_j}{cp_j}\right)
                      A_{i_1 j}B_{j i_2}\right)^2 }   .
\end{eqnarray}
By multiplying out the right hand side, it follows that
\begin{eqnarray}
\hspace{-0.2in}
\Expect{\FNormS{AB - CR}} 
\hspace{-0.1in}
\nonumber &=& \Expect{\sum_{i_1=1}^m \sum_{i_2=1}^p 
                      \sum_{j_1 \in \Lambda}\sum_{j_2 \in \Lambda}
                      \left(1-\frac{I_{j_1}}{cp_{j_1}}\right)
                      \left(1-\frac{I_{j_2}}{cp_{j_2}}\right) 
                      A_{i_1 j_1}B_{j_1 i_2}
                      A_{i_1j_2}B_{j_2 i_2} }   \\
\hspace{-0.3in}
\label{eqn:lab2}
          &=& \sum_{i_1=1}^m \sum_{i_2=1}^p 
              \sum_{j_1 \in \Lambda}\sum_{j_2 \in \Lambda}
              \Expect{\left(1-\frac{I_{j_1}}{cp_{j_1}}\right)
                      \left(1-\frac{I_{j_2}}{cp_{j_2}}\right)} 
              A_{i_1 j_1}B_{j_1 i_2} A_{i_1 j_2}B_{j_2 i_2}   .
\end{eqnarray}
Notice that for $j \in [\Lambda]$, $\Expect{1-I_{j}/cp_{j}}=0$ and 
$\Expect{\left(1-I_{j}/cp_{j}\right)^2}=\left(1/cp_j\right)-1 \leq 1/cp_j$.
Hence, 
\begin{eqnarray}
\nonumber 
\Expect{\FNormS{AB - CR}} 
   &=&    \sum_{i_1=1}^m \sum_{i_2=1}^p \sum_{j \in \Lambda}
          \Expect{\left(1-I_{j}/cp_{j}\right)^2} A_{i_1 j}^2 B_{j i_2}^2   \\
\nonumber
   &\leq& \sum_{j \in \Lambda} \frac{1}{cp_j}
          \sum_{i_1=1}^m \sum_{i_2=1}^p A_{i_1 j}^2 B_{j i_2}^2
    =     \frac{1}{c} \sum_{j \in \Lambda} 
                      \frac{\VTTNormS{A^{(j)}}\VTTNormS{B_{(j)}}}{p_j}   .
\end{eqnarray}
This concludes the proof of (\ref{eqn1:prf:thm:matmult-main-expected}) and 
thus of the theorem.
\end{Proof}


\end{document}